\documentclass[letterpaper,11pt]{article}
\usepackage{jheppub}
\usepackage{amsmath}
\usepackage{bm}
\usepackage{amssymb}
\usepackage{listings}
\usepackage{multirow}
\usepackage{makecell}
\usepackage{array}
\usepackage{subfigure}
\usepackage{siunitx}
\usepackage{booktabs}
\usepackage{amssymb}
\usepackage{pifont}
\usepackage{listings}
\usepackage{xcolor}
\usepackage{float}
\usepackage{slashed}

\usepackage{tikz}
\usepackage{pgffor}							

\newcommand{\be}{\begin{equation}}
\newcommand{\ee}{\end{equation}}
\newcommand{\beq}{\begin{equation}}
\newcommand{\eeq}{\end{equation}}
\newcommand{\bea}{\begin{eqnarray}}
\newcommand{\eea}{\end{eqnarray}}
\newcommand{\besp}{\begin{equation}\begin{split}}
\newcommand{\eesp}{\end{split}\end{equation}}

\newcommand{\Dfbd}{\mathord{\buildrel{\lower3pt\hbox{$\scriptscriptstyle\leftrightarrow$}}\over {D}_{\mu}}}

\def\0{\textbf{0}}
\def\1{\textbf{1}}
\def\2{\textbf{2}}
\def\3{\textbf{3}}
\def\4{\textbf{4}}
\def\5{\textbf{5}}
\def\6{\textbf{6}}
\def\7{\textbf{7}}
\def\8{\textbf{8}}
\def\9{\textbf{9}}

\begin{document}

\title{Scattering of non-relativistic finite-size particles and puffy dark matter direct detection}

\author[1]{Wu-Long Xu}
\author[2,3]{,~Jin Min Yang}
\author[4]{,~Jun Zhao}

\affiliation[1]{School of Electronic Engineering, Chengdu Technological University, Chengdu 611730, P. R. China}
\affiliation[2]{Center for Theoretical Physics, Henan Normal University, Xinxiang 453007, P. R. China}
\affiliation[3]{Institute of Theoretical Physics, Chinese Academy of Sciences, Beijing 100190, P. R. China}
\affiliation[4]{School of Physics and Electronics, Henan University, Kaifeng 475004,  P. R. China}

\emailAdd{396440567@qq.com}
\emailAdd{jmyang@itp.ac.cn}
\emailAdd{junzhao@henu.edu.cn}

\abstract{
In this work we consider the scattering between non-relativistic particles with different finite sizes. We first  calculate their interaction potential and apply the partial wave method to obtain their scattering cross section. Our findings show that the particle size can significantly affect the scattering between non-relativistic particles. Then we apply such a study to direct detection of puffy dark matter. We find that the finite size of the target nucleus may introduce non-perturbative effects that differ from the scenario of point-like dark matter. For large-size dark matter particles, this non-perturbative regime in the dark matter–nucleus scattering cross section effectively disappears; while for small values of the size-to-range ratio in the scattering process, a significant non-perturbative regime can maintain. Finally, for the direct detection of nugget-type puffy dark matter with a small number of constituent particles, we find that the stability conditions for the formation of bound-state dark matter can provide constraints on the  dark matter–nucleus scattering cross section.}

\maketitle


\section{Introduction}\label{sec1}
Dark matter dominates the matter content of the universe~\cite{Bertone:2016nfn,Tian:2020tur}, with evidence primarily coming from gravitational interaction, such as the galaxy rotation curves, the bullet cluster collision, and the observation of the cosmic microwave background (CMB)~\cite{Ostriker:1973uit,Randall:2008ppe}. The dark matter particle candidates can have a mass in a vast range, spanning from $10^{-22}\rm eV$ to several tens of the solar mass~\cite{Lin:2019uvt}, with the most popular one being the Weakly Interacting Massive Particle (WIMP) because it can naturally provide the relic abundance and is accessible at collider experiments~\cite{Buchmueller:2017qhf,Jungman:1995df,Lee:1977ua}. Unfortunately,  so far no unambiguous WIMP evidence  has been observed in laboratory searches~\cite{GAMBIT:2017zdo,XENON:2024wpa,XENON:2025vwd}, which makes a light or ultralight dark matter candidate increasingly attractive. Since lighter dark matter particles give lower recoil energies in scattering off nucleus, they are much harder to detect in laboratory experiments and thus more sensitive instruments or advanced experimental techniques are needed. 

For direct detection of light dark matter, several approaches have been considered: dark matter–electron scattering~\cite{Essig:2017kqs,XENON:2024znc}, the Migdal effect~\cite{Knapen:2020aky}, cosmic ray–boosted dark matter (CRDM)~\cite{Wang:2021jic,LZ:2025iaw}, levitated particle detectors~\cite{Kilian:2024fsg,Cheng:2023loy}, and varying the target material from noble gases to superconductors or semiconductors~\cite{Liang:2022xbu,Hochberg:2016ajh,Liang:2024xcx,Kurinsky:2019pgb}. Some strategies for detecting ultralight dark matter have also been proposed \cite{An:2024wmc}. Anyway, for direct detection of light dark matter, an accurate calculation of the dark matter–nucleus scattering cross section is crucial.  Typically, the traditional scattering cross section is modeled as a constant prefactor multiplied by a momentum-transfer-dependent term and a form factor that accounts for the finite size of the target nucleus~\cite{Acevedo:2021kly,Hardy:2015boa,Bhoonah:2020dzs,Cappiello:2020lbk}. However, in the non-relativistic regime, non-perturbative effects in particle scattering should be taken into account in order to obtain a realistic cross section~\cite{Bollig:2021psb,Bollig:2024ipe}. A notable example is the inclusion of non-perturbative effects in point-like dark matter self-scattering to address the small-scale structure problems in astronomy \cite{Tulin:2013teo}. Furthermore, at low velocities, attractive interactions between particles may lead to bound-state formation. Non-perturbative effects can also significantly enhance the annihilation cross section of low-energy point-like particles through the Sommerfeld enhancement mechanism~\cite{Lee:1977ua,Beneke:2022rjv}. On the other hand, the finite-size effect of particles may also play a critical role in their scattering. When considering the scattering of finite-size dark matter particles off target nuclei, an additional form factor associated with the size of the dark matter must be included, which leads to a larger cross section exclusion region compared to point-like dark matter \cite{Wang:2023xgm}. This remains true even in direct detection of semi-relativistic dark matter like the CRDM \cite{Wang:2023wrx}. Actually, the particle size effect may directly influence its low-energy scattering, e.g., Ref. \cite{Xu:2020qjk} discussed how the nuclear radius affects the dark matter-nucleus scattering cross section, Ref. \cite{Wang:2023xii} showed that the finite-size dark matter self-scattering cross section can be categorized into Born, resonant, and classical regimes due to size effects, while Ref. \cite{Wang:2023wbw} examined the influence of size effects on the Sommerfeld enhancement in the annihilation cross section of finite-size dark matter.

Therefore, both non-perturbative and finite-size effects should be considered when analyzing scattering between non-relativistic, finite-size particles~\cite{Digman:2019wdm}.
In this work, we will study the scattering between different finite-size particles and apply this size effect to the calculation of scattering cross section in the direct detection of puffy dark matter. First, we will re-examine the interaction potential between finite-size particles. Traditionally, such a potential is modeled as the point-like interaction potential multiplied by a form factor that captures the finite-size effect~\cite{Chu:2018faw}. Consequently, the scattering cross section is often given by the tree-level quantum field theory result for point particles multiplied by a form factor, namely, $d\sigma/d\Omega=\left(d\sigma/d\Omega\right)_{\rm point-like}F_{\rm size}(q)$, where $q$ is the momentum transfer~\cite{Gelmini:2002ez}. This result is equivalent to that obtained from quantum mechanical methods~\cite{Wang:2021tjf}. However, in the non-relativistic regime, a more precise treatment requires using quantum mechanics in coordinate space, integrating over the spatial extent of the finite-size target nucleus to obtain the correct dark matter–nucleus interaction potential. We will perform a state-of-the-art computation that incorporates the particle size effect accurately. Then by solving the Schrödinger equation, we will obtain a more realistic scattering cross section between finite-size particles, which includes non-perturbative effects in the non-relativistic regime~\cite{Xu:2020qjk}.

This paper is organized as follows. In Sec.\ref{sec2}, we present our treatment for the scattering between different non-relativistic finite-size particles, including the modified Yukawa potential, the conditions for the Born approximation, and the partial wave method. In Sec.\ref{sec3}, we apply our calculation to the direct detection of puffy dark matter. We first consider the impact of the target nucleus size on the scattering cross section in the direct detection of point-like dark matter; then we investigate the size effects in the scattering between finite-size dark matter and nucleons; finally, based on the stability conditions of nugget-type puffy dark matter particles, we present the parameter space of dark matter mass and scattering cross section relevant for direct detection.  Sec. \ref{sec4} gives our conclusions.

\section{Scattering of non-relativistic finite-size particles}\label{sec2}
In this section, we consider the scattering of non-relativistic finite-size particles.
While some part of the content is a re-description of the knowledge in the literature, 
we will calculate the modified Yukawa potential between finite-sized particles
in order to consider the non-perturbative effects in the scattering. 
We will also solve the Schrödinger equation using the potential in spatial coordinates to obtain the scattering cross-section. 

Theoretically, the Compton wavelength characterizes the quantum fluctuation scale of a particle at rest and is related to its positional uncertainty. If a particle's size exceeds its Compton wavelength, its position can be localized, which is a necessary condition for the existence of a finite-size particle. On the other hand, the de Broglie wavelength describes the wave-like nature of a moving particle. When this wavelength is larger than the size of the target particle being probed, the moving (probing)  particle perceives the target as a point-like particle. Conversely, if the de Broglie wavelength of the moving particle is smaller than the size of the target particle, the internal structure of the target can be resolved~\cite{Colquhoun:2020adl}.

Similarly, for non-relativistic scattering between particles, we must consider the effective range of interaction (also called the force range), which is typically characterized by the inverse of the mediator mass, $1/m_{\phi}$. When the force range is much smaller than the incident particle’s de Broglie wavelength (
$1/m_{\phi}\ll1/mv$), the interaction must be treated quantum mechanically, and the scattering is dominated by the s-wave. When the force range is comparable to the de Broglie wavelength ($1/m_{\phi}\sim1/mv$), resonant or bound-state phenomena may occur during scattering~\cite{Grabowska:2018lnd}. When the force range is much larger than the de Broglie wavelength, the scattering corresponds to a long-range interaction and the wave nature of the incident particle no longer affects the spatial distribution of the force, in which the interaction can be treated semiclassically and the field-theoretic methods are typically used to calculate the scattering cross section.

If a non-relativistic particle has finite size, then, as discussed in Ref. \cite{Wang:2023xii}, the interaction potential between two such finite-size particles exhibits a sharp cutoff beyond their spatial extent due to the size effect. This short-range behavior is analogous to (albeit distinct from) nuclear forces (here we only consider elastic scattering). For traditional long-range potentials such as Coulomb or Yukawa interactions, the short-range size effect fundamentally modifies the calculation of the scattering cross section between finite-size particles. Therefore, when dealing with scattering between different finite-size particles, one must reconsider the key parameters used in classifying scattering cross sections, rather than directly applying the classification used for point-like non-relativistic particles.

Here, we introduce some conventions for finite-size particles (also called puffy particles).
Take finite-size dark matter particle as an example: a bound-state dark matter particle is composed of N point-like dark matter constituents, each with mass $m_{\chi}$, resulting in a total mass of $\rm Nm_{\chi}$~\footnote{Here we neglect the binding energy and do not consider any dark strong interaction between the constituents. However,  the scattering may be changed enormously by dark strong interaction  in some parameter space~\cite{Wang:2021tjf}.}.

\subsection{The modified Yukawa  potential}\label{sec2.1}
The interaction between particles affects their scattering cross section. Here, we recalculate the interaction potential between low-velocity, finite-size particles in coordinate space. We begin with the Yukawa interaction potential between two point-like fermions, which is expressed as 
\be \label{eq1}
V(r)=-\frac{\alpha}{r} e^{-m_{\phi}r},
\ee
with $m_{\phi}$ representing the mass of the mediator and $r$ being the distance between the two particles.  The  fine structure constant is conventionally defined as $\alpha = g^2/4\pi$ with $g$ being the coupling constant. The interaction potential between two finite-volume particles is expressed in terms of their charge density functions as
\be \label{eq2}
V(r)=\frac{1}{4\pi}\int dV_1 dV_2
\rho_{1}(r_1)\frac{e^{-m_{\phi}
		|r_1-r_2|}}{
	|r_1-r_2|}\rho_{2}(r_2)\, ,
\ee
where $\rho_{1}(r_1) $ and $\rho_{2}(r_2)$ are the charge density functions, and $r$ is the distance between the two center points of the particle charges (in our calculation we choose the charge center point of one particle like a nucleus or a nucleon as the origin of coordinates).   
In general, the charge density function of a particle can take various forms, such as dipole, tophat, and Gaussian. In Ref.~\cite{Chu:2018faw}, the Hamiltonian between finite-size particles was computed in momentum space and found to be largely insensitive to the specific form of the charge density function. Similarly, Ref.~\cite{Wang:2023xii} also showed that the impact of the charge density profile on the potential is negligible. Therefore, in this work, we consider only spherically symmetric finite-size particles with a tophat-type charge density, given by 
\be 
\rho(r) = \frac{3}{4\pi R_0^3}\theta(R_0 - r), 
\ee
where $R_0$ is the radius of the finite-size particle. When one of the two particles is point-like, Eq.~(\ref{eq2}) reduces to
\be \label{eq3}
V(r)=\frac{1}{4\pi}\int dV_1 
\rho_{1}(r_1)\frac{e^{-m_{\phi}
		|r-r_1|}}{
	|r-r_1|}\, .
\ee
And when the other finite-size particle has a tophat charge density, their interaction  potential  can be written as 
\begin{align}\label{eq4}
V_{\rm point-N}(r) & = \begin{cases}
~\frac{3\alpha}{R^3_0}\times \left(-\frac{e^{-m_{\phi}(r+R_0)}(e^{2m_{\phi}r-1})(1+m_{\phi}R_0)}{2m_{\phi}^3r}+\frac{1}{m_{\phi}^2}\right) & {\rm for}~ r<R_{0}\, , \\
\hspace{5cm}\  & \ \\[-6.mm]
~\frac{3\alpha e^{-m_{\phi}r}}{2m_{\phi}R^3_0r}\left(\frac{e^{-m_{\phi}R_0}}{m_{\phi}^{2}}-\frac{e^{m_{\phi}R_0}}{m_{\phi}^{2}}+\frac{R_0e^{-m_{\phi}R_0}}{m_{\phi}}+\frac{R_0e^{m_{\phi}R_0}}{m_{\phi}}\right)&  {\rm for}~  r>R_{0}\,.
\end{cases}
\end{align}
\begin{figure}[h]
	\centering
	\includegraphics[width=10.cm]{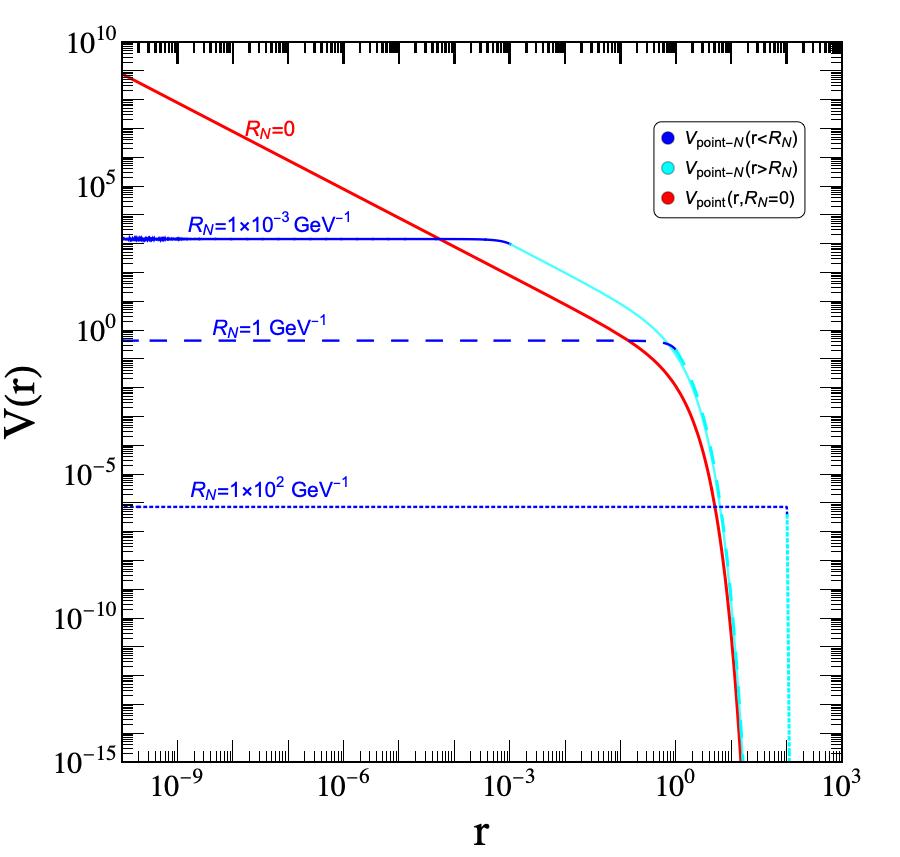}	
    \vspace{-.3cm} 
	\caption{The interaction potential between two particles: the red line represents the Yukawa potential between two point particles, the blue lines (for $r<R_{\rm N}$) and the cyan-green lines (for $r>R_{\rm N}$) represent the modified Yukawa potential between a point-like dark matter particle and a finite-size target nucleus with a radius $R_N$. }
	\label{fig1}
\end{figure}

Fig.~\ref{fig1} illustrates the behavior of the interaction potential between a point-like particle and a finite-size particle. The red solid line represents the conventional Yukawa potential between point particles, which diverges as the distance $r$ approaches zero. In contrast, the blue solid, dashed, and dotted lines represent the potentials between a point particle and a finite-size particle with different radii. These curves show that, due to finite-size effects, the potential is essentially confined within the radius of the finite-size  particle and drops sharply beyond the radius. For a small-radius finite-size particle, the potential within the radius tends to be nearly constant, and beyond the radius it decreases gradually before dropping off sharply, as shown by the blue-green solid and dashed lines in Fig.~\ref{fig1}. This demonstrates that the presence of size effects leads to a fundamentally different interaction behavior compared with point particles. In the case of a large-radius particle, the interaction (similar to nuclear forces) is confined within the radius and remains constant in magnitude.
Next, we consider Eq.~(\ref{eq2}) with both colliding particles being finite-size and having a tophat charge density. In this case, the explicit expression for the interaction potential becomes
\begin{align}\label{eq5}
V_{\rm \chi-N}(r) & = \begin{cases}
~ g(r,R_{\chi},R_N) &  {\rm for}~  r<2R_{\chi}\, , \\
\hspace{5cm}\  & \ \\[-6.mm]
~\alpha\frac{e^{-m_{\phi}r}}{r} 
 h\left(R_{\chi},R_N\right)&   {\rm for}~  r>2R_{\chi}\,,
\end{cases}
\end{align}
with the functions $g(r,R_{\chi},R_N)$ and $h\left(R_{\chi},R_N\right)$ given in Appendix \ref{appa}. Since we will 
study the puffy dark matter direct detection in the proceeding section, here we assume one of the particles to be a puffy dark matter particle with a radius $R_{\chi}$. 

\begin{figure}[h]
\centering
\includegraphics[width=10.cm]{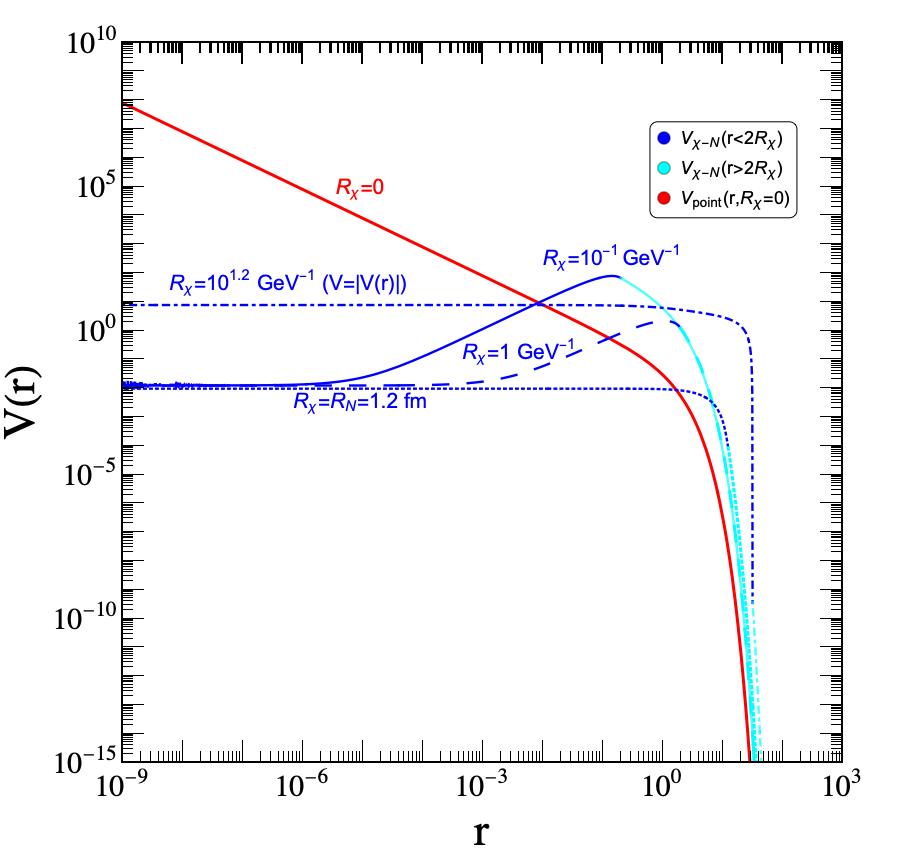}	
    \vspace{-.3cm} 
\caption{Interaction potential between particles: the red line represents the Yukawa potential between two point particles, the blue lines (for $r<2 R_{\chi}$) and the cyan-green lines (for $r>2R_{\chi}$) represent the modified Yukawa potential between a puffy dark matter particle and a finite-size target nucleus for different $R_{\chi}$ values. }\label{fig2}
\end{figure}

To demonstrate  the size effects in the scattering between two particles, we take the dark matter-nucleon scattering as an example. We  fix the nucleon size a typical value $R_N=1.2~\rm fm\sim6.2~GeV^{-1}$ and compare several cases with different dark matter radii.
As shown in Fig.~\ref{fig2}, in contrast to the Yukawa potential between point particles, the interaction potential between two finite-size particles tends to be a constant as $r$ gets very small.  For a puffy dark matter particle with a size comparable to or larger than the nucleon size, the potential remains nearly a constant before dropping sharply. While for a dark matter particle size smaller than the nucleon size, the potential first remains flat, then increases slightly and finally drops off as $r$ increases.  This behavior deviates from the point dark matter particle–nucleon interaction. So the dark matter-nucleon scattering cross section will also differ from the point-like case. 

\subsection{Born approximation condition}\label{sec2.2}
When the kinetic energy is much larger than the potential energy, the quantum mechanical scattering cross section coincides with the tree-level cross section calculated in quantum field theory, which is commonly used in dark matter direct detection. Therefore, we can use Born approximation in quantum mechanics to estimate the scattering cross section in the presence of finite-size effects. From the above potential calculation we see that in the case of large-size particles or the two particles with equal sizes, the modified interaction potential between the particles is very small and can be approximated as a piecewise function:
\begin{align}\label{eq6}
V(r) & = \begin{cases}
~ V_0 & {\rm for} ~r<R_{N}\, , \\
\hspace{1cm}\  & \ \\[-6.mm]
~0&  {\rm for} ~r>R_{N}\,.
\end{cases}
\end{align}
Under Born approximation, the scattering amplitude is given by
\be\label{eq7}
f^1(q)=-\frac{2\mu_A}{q}\int^{\inf}_0V(r')\sin(qr')r'dr',
\ee
where $q=2k\sin\theta/2$ is the momentum transfer and $\mu_A$ is the reduced mass. Substituting the potential into the scattering amplitude yields:
\be\label{eq8}
f^1(q)=\frac{2\mu_AV_0}{q^3}\left[qR_N\cos(qR_N)-\sin(qR_N)  \right].
\ee
Therefore, under the first-order Born approximation,
\be\label{eq9}
2\mu_{A}\left|\int_{0}^{\infty}rV(r)dr\right|\ll1
\, .
\ee
The total scross section becomes
\be\label{eq10}
\sigma_{A}=\frac{\pi \mu_A^2V_0^2}{16k^6}\left[4kR_N\sin(4kR_N)+\cos(4kR_N)+32k^4R^4_N-8k^2R^2_N-1 \right].
\ee
In the limit $kR_N\ll1$, it can be simplified to
\be\label{eq11}
\sigma_{A}\approx\frac{16\pi}{9}\mu_A^2R_N^6V_0^2.
\ee

Under the Born approximation condition $\mu_AR^2_NV_0\ll1$, the result can be rewritten as $\sigma_A\ll16\pi/9R^2_N$. This shows that the scattering cross section is closely related to the size of the particle. In dark matter detection, taking $R_N\approx1.2~\rm fm$, the Born approximation sets an upper limit on the scattering cross section $\sigma_A\ll10^{-25}\rm cm^2$.

However, for non-relativistic low-energy or small-size particles, calculating only the s-wave contribution is not accurate. To assess the deviation between the Born approximation in quantum field theory and the actual scattering cross section affected by finite-size effects, we will investigate the conditions under which the Born approximation is valid for scattering between finite-size particles. This will then be compared with the original applicable parameter space for the Born approximation of point particles. For the point-like Yukawa potential, the Born approximation condition is $2\mu_A\alpha/m_{\phi}\ll1$. First, we consider the Born approximation condition for scattering between a point particle and a finite-size particle. By substituting the potential of Eq.(\ref{eq4}) into Eq.(\ref{eq9}), we obtain 
\begin{eqnarray}
2\mu_A
\left|\int_{0}^{\infty}rV(r)dr\right|
= \frac{2\mu_A\alpha}{m_{\phi}}
\left|\frac{3e^{-y}(2+2y+(y^2-2)e^y)}{2y^3}\right|
=bf(y) \ll 1\,,
\label{eq12}
\end{eqnarray}
where we define the ratio of the finite-size particle’s radius to the force range as $y=m_{\phi}R_N$ and the dimensionless parameter $b=2\mu_A\alpha/m_{\phi}$.

\begin{figure}[ht]
	\centering
	\includegraphics[width=10.cm]{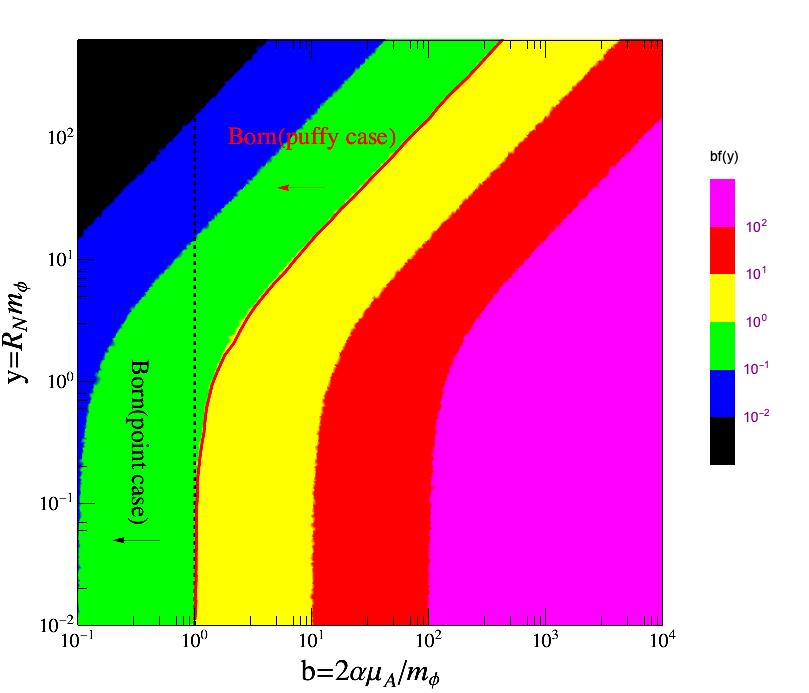}	
    \vspace{-.3cm} 
	\caption{The parameter space of $y$ versus $b$ for different values of $bf(y)$, showing the Born and nonperturbative regimes separated by the black dashed line for point-like particles or the solid red curve for the finite-size particles.}
	\label{fig3}
\end{figure}

Fig.~\ref{fig3} shows the contour plot of the Born approximation parameter $b$ versus the radius-to-range ratio $y$ for scattering between a point particle and a finite-size particle. We see that the Born approximation region for point-particle scattering lies to the left of the black dashed line, but when finite-size effects are included, this region extends to the left of the red solid line. This means that some parameter regions that originally did not satisfy the Born approximation for point-particle scattering may fall within the Born approximation region once finite-size effects are taken into account. Therefore, for non-relativistic scattering between finite-size particles, it is necessary to reconsider the solution of the Schrödinger equation with the modified potential to obtain the true scattering cross section.

\begin{figure}[ht]
\centering
\includegraphics[width=5.2cm]{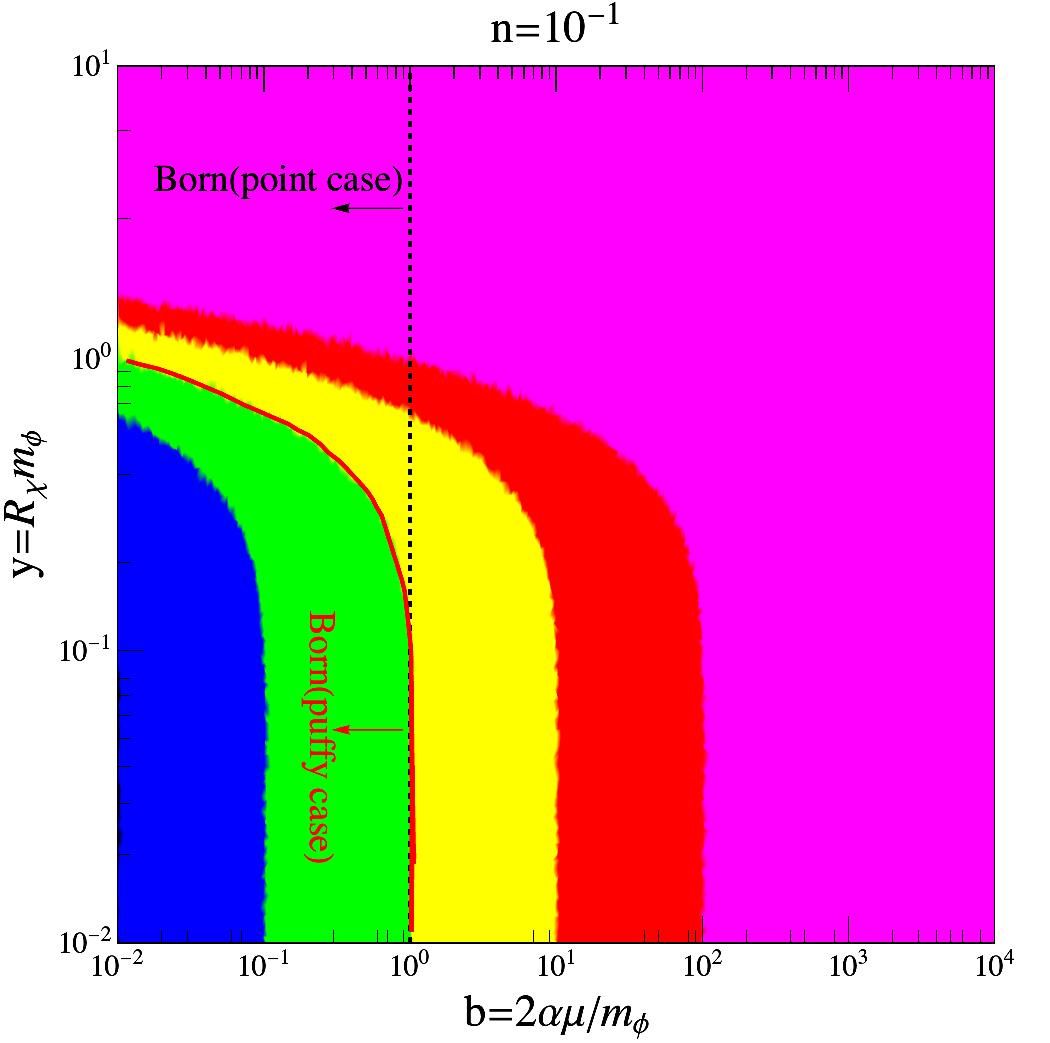}\hspace{-7mm}	
\includegraphics[width=5.2cm]{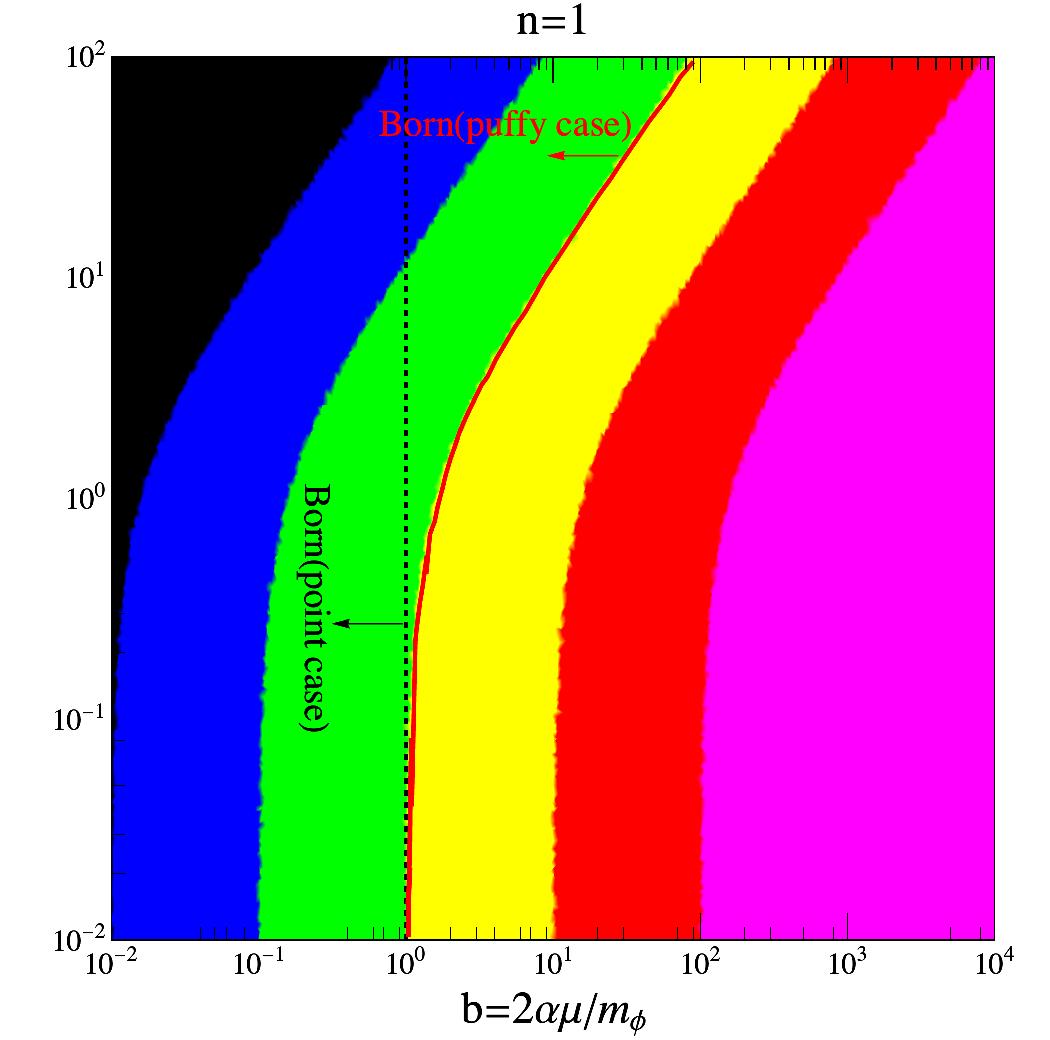}\hspace{-7mm}	
\includegraphics[width=6.2cm]{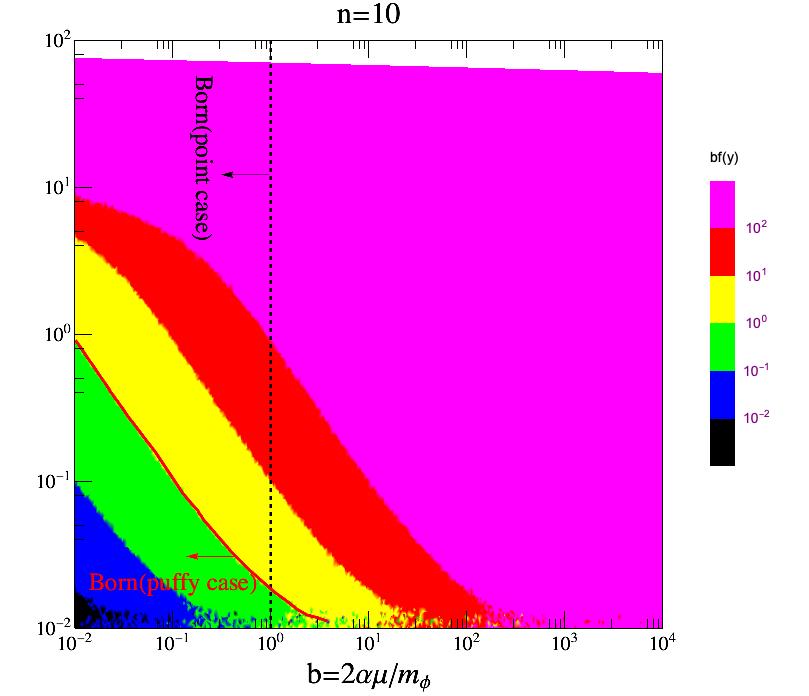}
\vspace{-.3cm} 
	\caption{Similar as Fig.\ref{fig3}, but for a finite-size dark matter particle scattering off a finite-size nucleon with different  $R_{\chi}/R_N=n$ values.}
	\label{fig4}
\end{figure}

Next, we consider the Born approximation conditions for scattering between particles of different sizes. As a specific example, we study the scattering between a nucleon with size 
$R_N=1.2~\rm fm$ and a dark matter particle with size $R_{\chi}=6.2n~\rm GeV^{-1}$, i.e., $R_{\chi}/R_N=n$. In this case, the interaction potential between the particles becomes
\begin{align}\label{eq13}
V_{\rm \chi-N}(r) & = \begin{cases}
~ g(r,6.2n~\rm GeV^{-1},6.2~\rm GeV^{-1}) & {\rm for}~ r<2R_{\chi}\, , \\
\hspace{5cm}\  & \ \\[-6.mm]
~\alpha\frac{e^{-m_{\phi}r}}{r} 
\times h(r,6.2n~\rm GeV^{-1},6.2~\rm GeV^{-1})&  {\rm for}~ r>2R_{\chi}\,.
\end{cases}
\end{align}
By substituting Eq.(\ref{eq13}) into the Born approximation condition formula Eq.(\ref{eq9}), we obtain the modified Born approximation condition for scattering between finite-size particles. Due to the complexity of the formula, here we do not present the detailed derivation; instead, the numerical results are shown in Fig.\ref{fig4} (note that different values of 
$n$ correspond to different $f(y)$ functions). This figure shows that the parameter space satisfying the Born approximation condition is sensitive to the sizes of the scattering particles. Comparing the different panels of Fig.\ref{fig4}, it is clear that when the size difference between the particles gets larger, the Born approximation region becomes smaller. This further demonstrates that using the tree-level calculation from quantum field theory to obtain scattering cross sections is not accurate in such a case.

\subsection{Partial wave approach}\label{sec2.3}
In this subsection, we recapitulate the calculation of the elastic scattering cross section between non-relativistic particles using the partial wave method. Since our primary focus is on direct detection of dark matter, where the momentum transfer is typically small when the scattering angle approaches zero in the detector, it is more useful to consider the momentum-transfer cross section:
\be\label{eq14}
\sigma_T=\int d\Omega(1-\cos\theta)d\sigma/d\Omega,
\ee 
with $\theta$ being the scattering angle. 
Beyond the Born approximation region, i.e., in the non-perturbative regime, there is no analytic formula for the scattering cross section and thus we can only perform numerical calculation using the partial wave method. Then we expand the scattering wave function in terms of Legendre polynomials to calculate the phase shift for each partial wave. The differential scattering cross section is given by 
\be \label{eq15}
\frac{d\sigma}{d\Omega}=\frac{1}{k^{2}}\Big|\sum_{l=0}^{\infty}(2l+1)e^{i\delta_{l}}P_{l}(\cos\theta)\sin\delta_{l}\Big|^{2},
\ee
with $k$ being the momentum of the incident particle while $\delta_l$ being the phase shift of the $l$-th partial wave. 
The phase shift $\delta_l$ can be obtained by solving the Schr\"odinger equation for the radial wave function $\mathcal{R}_l(r)$: 
\be\label{eq16}
\frac{1}{r^{2}}\frac{\partial}{\partial r}\left(r^{2}\frac{\partial\mathcal{R}_{l}}{\partial r}\right)+\left(E-V(r)-2\mu V(r)\right)\mathcal{R}_{l}(r)=0.
\ee
From the asymptotic solution for $\mathcal{R}_l(r)$ we have   
\be \label{eq17}
\underset{r\rightarrow\infty}{\lim}\mathcal{R}_{l}(r)\propto\cos\delta_{l}j_{l}(kr)-\sin\delta_{l}n_{l}(kr),
\ee
with $j_l$ being the spherical Bessel function and $n_l$ being the spherical Neumann function. In terms of the phase shift, the transfer cross-section is given by 
\be\label{eq18}
\frac{\sigma_{T}k^{2}}{4\pi}=\sum_{l=0}^{\infty}(l+1)\sin^{2}(\delta_{l+1}-\delta_{l}).
\ee
With the definition of some dimensionless parameters
\be \label{eq19}
\chi_l=rR_l, \quad  x=2\alpha \mu r, \quad  a=\frac{v}{2\alpha},\quad  b=\frac{2\mu \alpha}{m_{\phi}}\, ,
\ee
the Schrödinger equation Eq.(\ref{eq16}) takes a form 
\be\label{eq20}
\left(\frac{d^{2}}{dx^{2}}+a^{2}-\frac{l(l+1)}{x^{2}}-\frac{1}{m_{\chi}\alpha^{2}}V(r)\right)\chi_{l}=0\, .
\ee
The details of solving the Schr\"odinger equation are presented in Appendix B of Ref. \cite{Wang:2023xii}. Here we just briefly describe the calculation.
The initial conditions like $\chi_l(x_i)=1$ and $\chi'_l(x_i)=(l+1)/x_i$ 
are set at a point $x_i$ near the origin. This leads to an angular momentum term dominating the Schrödinger equation, which is then solved within the range $x_i \leq x\leq x_m$, with $x_m$ being the maximum value of $x$ used in the numerical analysis. With the condition of  asymptotic solution  Eq.~(\ref{eq17}) and $x=x_m$, we have 
\be\label{eq21}
\chi_l\propto x e^{i\delta_{l}}\left[\cos\delta_lj_l(ax)-\sin\delta_ln_l(ax)\right].
\ee
Then we obtain 
\be\label{eq22}
\tan\delta_l=\frac{ax_mj'_l(ax_m)-\beta_lj_l(ax_m)}{ax_mn'_l(ax_m)-\beta_ln_l(ax_m)},
\ee
with $\beta_l$ given by 
\be
\beta_l=\frac{x_m\chi'_l(x_m)}{\chi_l(x_m)}-1.
\ee
Finally, by substituting the phase shifts into Eq.~(\ref{eq18}), we obtain the momentum-transfer scattering cross section. For non-relativistic self-scattering between point particles, the partial wave method reveals that the self-scattering parameter space is divided into Born, resonance, and classical regimes \cite{Tulin:2013teo}. The non-perturbative dynamics in the resonance and classical regimes can address the small-scale structure anomalies in cosmology. Similarly, when applying the partial wave method to calculate the self-scattering cross section of finite-size dark matter, it is found that contributions from partial waves with $l>0$ are non-negligible, and the cross section obtained solely from the Born approximation does not represent the true value.

\section{Size effect in puffy dark matter detection}\label{sec3}
The scattering cross section between a finite-size dark matter particle and a target nucleus is often assumed to be proportional to the square of the number of constituent particles comprising the dark matter bound state multiplied by the point particle–nucleus scattering cross section. However, this does not consider the volume effect of the finite-size dark matter particle. As can be seen from the modified potential in Eq.(\ref{eq5}) and Fig.~\ref{fig2}, the size effect has a significant impact on the scattering. Therefore, we should start from the fundamental interaction and perform a volume integration for finite-size dark matter particle to obtain a correct modified Yukawa potential, and then provide a more accurate scattering cross section for puffy dark matter detection.   We will first re-calculate the scattering cross section between a point dark matter particle and a finite-size nucleus. We reproduced the results in Ref.~\cite{Xu:2020qjk}, which showed that such a cross section can be divided into Bonn, resonance and classical regions. Then we will extend the study of this scattering to the case of two finite-size particles, i.e., a finite-size dark matter particle and a finite-size nucleus. Finally, we will provide the results for the nugget dark matter particle.

\subsection{Point dark matter-nucleus scattering}\label{sec3.1}
To intuitively illustrate the impact of particle size effect on scattering, we first show the finite-size effect of a target nucleus in point dark matter-nucleus scattering.  Traditionally, the scattering between a point dark matter particle and a target nucleus is calculated by multiplying the tree-level scattering amplitude from quantum field theory with a form factor that describes the nucleus charge density distribution. To take into account the non-relativistic effect of the dark matter scattering and the finite size effect of the target nucleus, we first compute the modified Yukawa potential between the point-like dark matter particle and the target nucleus in coordinate space, i.e., Eq.(\ref{eq4}) (for simplicity, here we consider the target nucleus as a proton with mass number $A=1$). Then, by substituting this potential into the Schrödinger equation Eq.(\ref{eq20}), we obtain phase shift solutions for different partial waves. Finally, the momentum-transfer scattering cross section can be calculated using Eq.~(\ref{eq18}).

\begin{figure}[ht]
	\centering
\includegraphics[width=5.5cm]{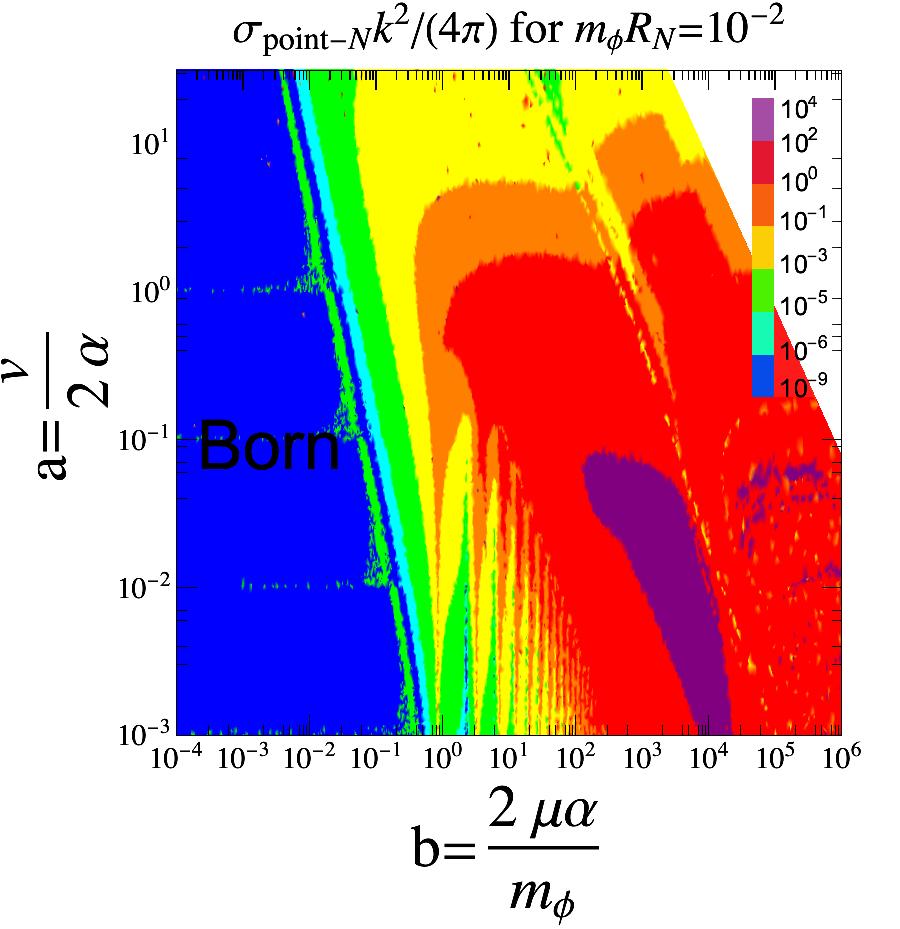}\hspace{-3.mm}
\includegraphics[width=4.9cm]{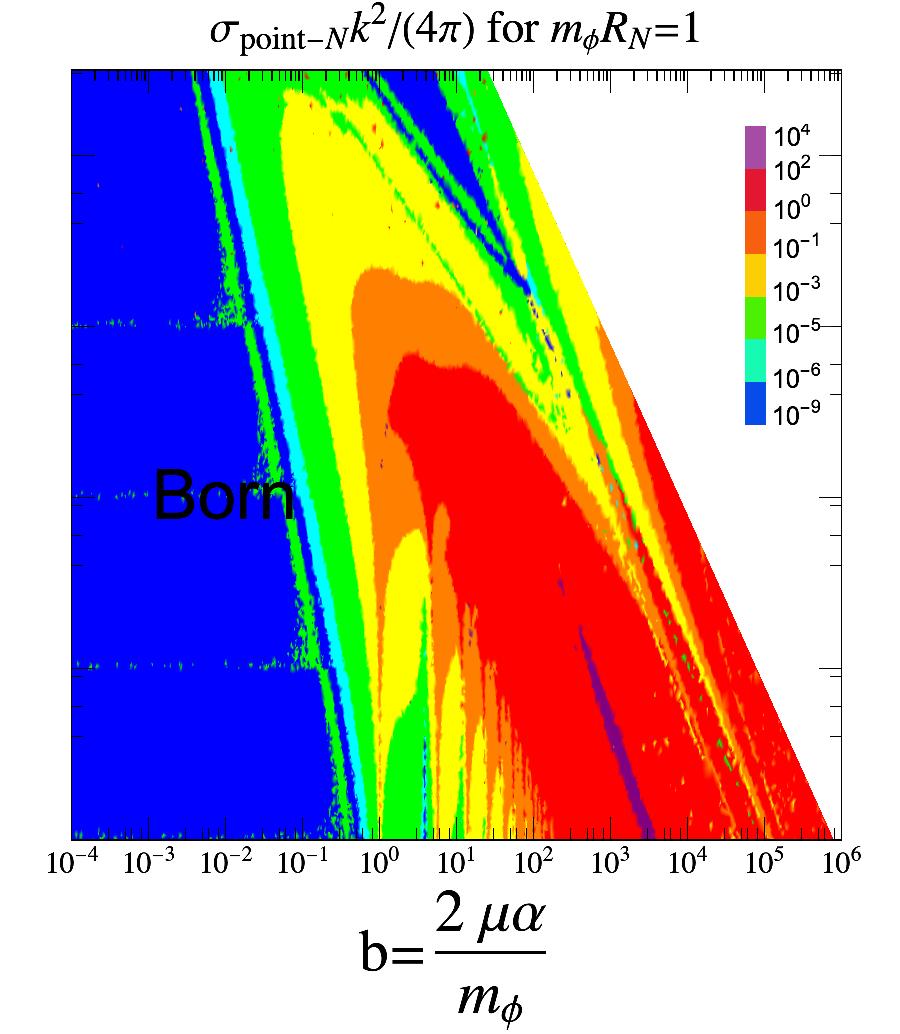}\hspace{-3.mm}
\includegraphics[width=4.9cm]{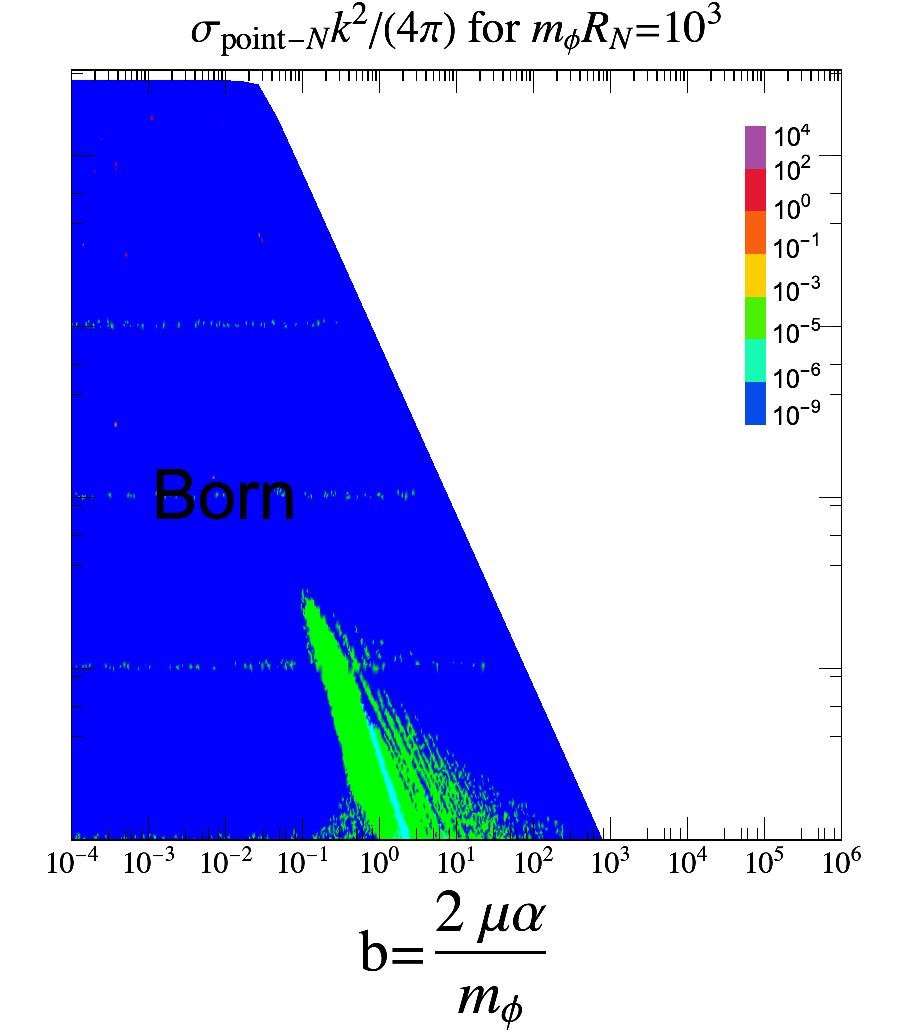}\\	
\vspace{-.3cm}
\caption{The parameter space of $a$ versus $b$ for point dark matter particle scattering off a proton with different $R_{\rm N}m_{\phi}$ values, showing the values of $\sigma_{\rm point-N}k^2/(4\pi)$. }\label{fig5}
\end{figure}
 
 We fix the dark matter velocity as 300 km/s, the proton mass as $m_N=0.938\rm~GeV$, and the reduced mass is defined as $\mu=(m_{\chi}m_N)/(m_{\chi}+m_N)$. By performing a random scan over the parameter space 
$(m_{\phi},\alpha, m_{\chi})$, we obtain the results shown in Fig.~\ref{fig5}. The right panel shows that for a large size-to-range ratio of the target nucleus, the scattering essentially lies within the Born approximation region. This is because in this regime the potential energy is nearly zero and much smaller than the kinetic energy, and consequently the scattering cross section is similar to the traditional point dark matter result. From the left and middle panels we see that for a small size-to-range ratio of the target nucleus, the scattering cross section classification resembles the result of point dark matter self-scattering, which is divided into Born, resonance, and classical regimes. Numerically, this arises from the competition between kinetic and potential energies of the scattering particles. In the s-wave case, the potential energy exceeds the kinetic energy, while as the partial wave number $l$ increases, the kinetic energy gradually dominates. However, unlike self-scattering, the competition between potential and kinetic energies here originates from the finite-size effect that modifies the Yukawa potential into a piecewise constant form rather than from the divergence behavior of the Yukawa potential itself. Even for a very small size-to-range ratio of the target nucleus, the size effect introduces a relatively large initial potential value in the piecewise potential. Furthermore, comparing the left and middle panels, we see that as the size-to-range ratio of the target nucleus increases, the parameter $a$ corresponding to the resonance peak decreases, until the resonance disappears entirely at a sufficiently large ratio. Similar results and analyses can also be found in Ref.~\cite{Xu:2020qjk}, which studied resonance scattering between point-like dark matter and baryons. 
 
\subsection{ Puffy dark matter-nucleus scattering}\label{sec3.2}
In this subsection, we calculate the scattering cross section between a finite-size dark matter particle and a target nucleus in direct detection. To study the impact of the number of nucleons in a target nucleus on direct detection, we calculate both puffy dark matter–proton scattering with nucleon number $A=1$ and puffy dark matter–xenon nucleus  scattering with nucleon number $A=132$. With the definition $R_{\chi}/R_N=n$, we have $R_{\chi}=6.2n\rm GeV^{-1}$ when the target is a proton and $R_{\chi}=30.67 n\rm GeV^{-1}$ when the target is a xenon nucleus. For puffy dark matter–proton scattering, the potential is given by Eq.(\ref{eq13}). Substituting this into the Schrödinger equation Eq.(\ref{eq20}), we obtain the phase shift solution for different partial waves. Then the momentum-transfer scattering cross section can be calculated using Eq.~(\ref{eq18}).

\begin{figure}[ht]
\centering
\includegraphics[width=7.2cm]{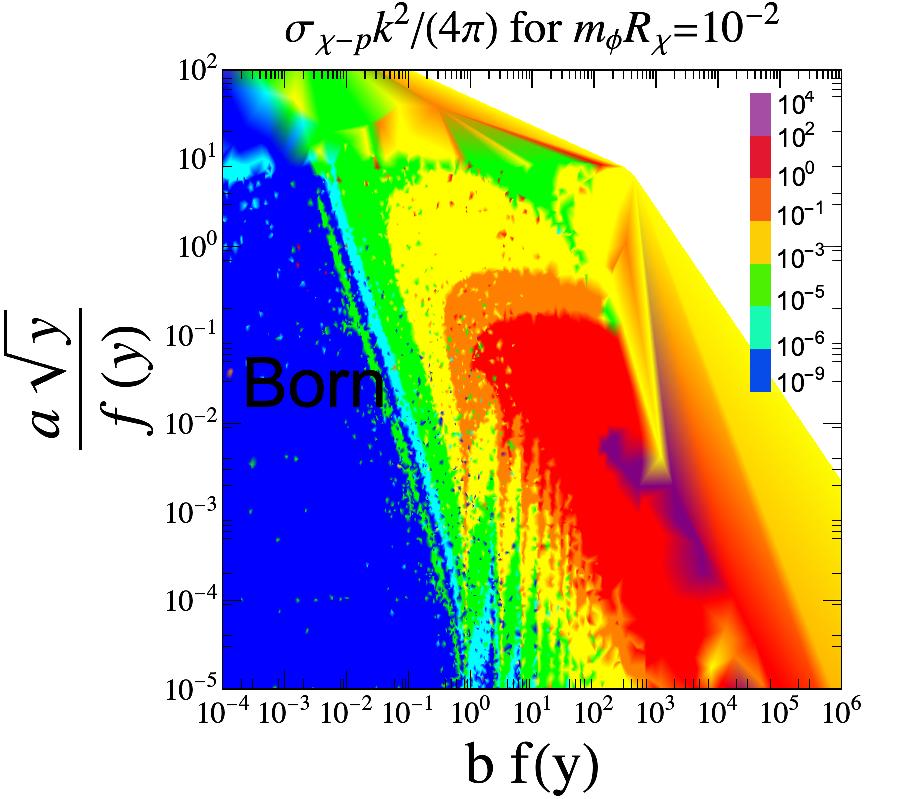}\hspace{-1mm}
\includegraphics[width=7.2cm]{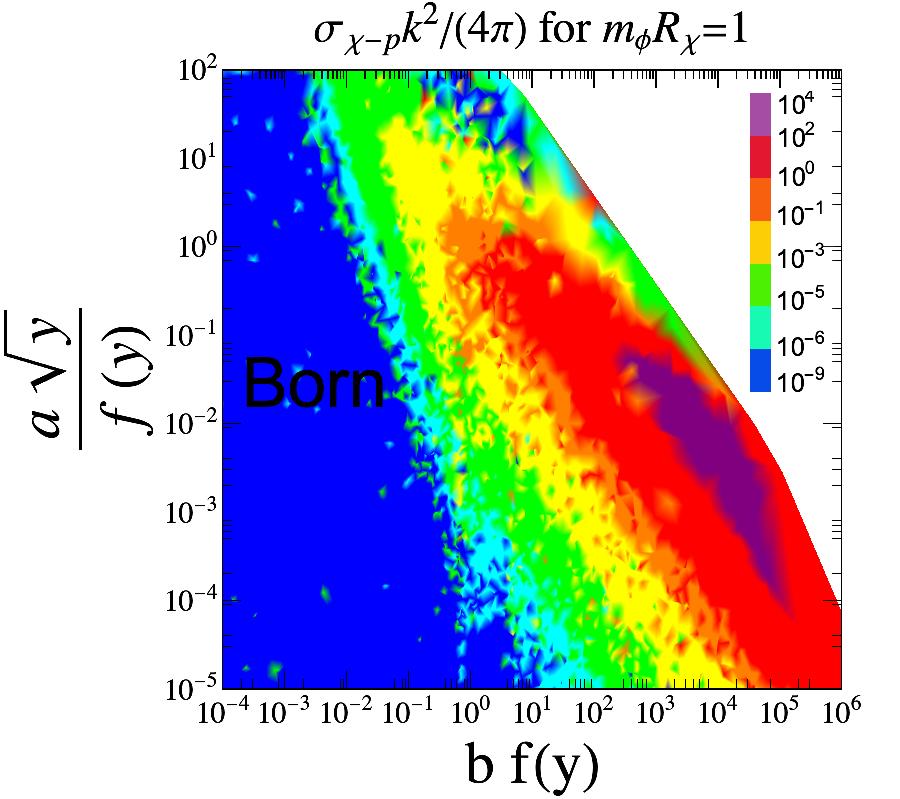}\hspace{-1mm}
    \vspace{-3mm}
\caption{The parameter space ($a\sqrt{y}/f(y), bf(y)$) for puffy dark matter-proton scattering with a fixed $R_{\chi}m_{\phi}$, showing the values of $\sigma_{\rm \chi-p}k^2/(4\pi)$. The $bf(y)<1$ region corresponds to Born approximation, while other regions are resonance or classical.  }\label{fig6}
\end{figure}

We define $y=R_{\chi}m_{\phi}$ and the dimensionless parameter $bf(y)$ can be obtained by substituting the potential Eq.(\ref{eq13}) into the Born approximation condition Eq.(\ref{eq9}) (due to the complexity of the expression, its explicit form is not presented here).
Then, we scan the parameter space ($m_{\phi}$, $\alpha$, $M_{\chi}$) with $M_{\chi}$ defined by $M_{\chi}=N m_{\chi}$ and obtain the results shown in Fig.~\ref{fig6}. Since we neglect the internal structure of the dark matter particle here, we consider only its total mass as a single parameter and do not specify the exact number of its constitutes. For a given value of $y$, we have 
$n=y/(m_{\phi}\times 6.2~\rm GeV^{-1})$. Fig.~\ref{fig6} displays the contour plots for the dimensionless parameters ($bf(y),a\sqrt{y}/f(y)$) with different scattering cross section values and  different $R_{\chi}m_{\phi}$ values. For larger $R_{\chi}m_{\phi}$ values such as $R_\chi m_{\phi}=10^{3}$, the vertical axis values approach zero and thus are not plotted here. The dark matter velocity is again taken as 300 km/s. Similar to point dark matter–nucleus scattering, appropriate parameter choice allows for the classification of scattering cross sections. For scattering between particles of different sizes, here we choose the horizontal axis as $bf(y)$ since it shows the Born approximation condition with values smaller than 1. The reason for choosing $a\sqrt{y}/f(y)$ as the vertical axis is explained in
Ref.\cite{Wang:2023xii}. 

From Fig.\ref{fig6} we observe that for small values of $y$, the scattering cross section can also be divided into Born, resonance, and classical regions. This is because the finite-size effect causes the interaction potential between scattering particles to start varying from a constant value. As the distance $r$ increases, the potential interaction varies depending on the ratio of the particle sizes, leading to different behaviors as shown in Fig.\ref{fig2}. Consequently, the competition between kinetic and potential energies inevitably arises, producing resonance and classical regions in the cross section. This complex non-perturbative dynamics influenced by particle size effects may give rise to rich phenomenology in particle scattering.

\begin{figure}[ht]
\centering
\includegraphics[width=7.2cm]{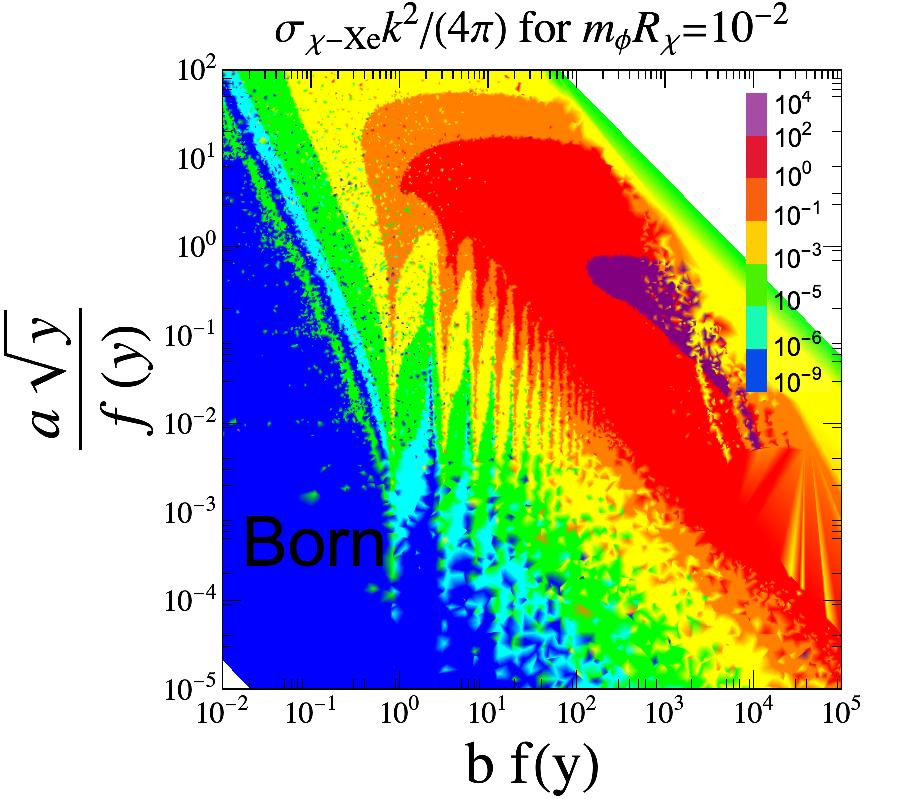}\hspace{-1mm}
\includegraphics[width=7.2cm]{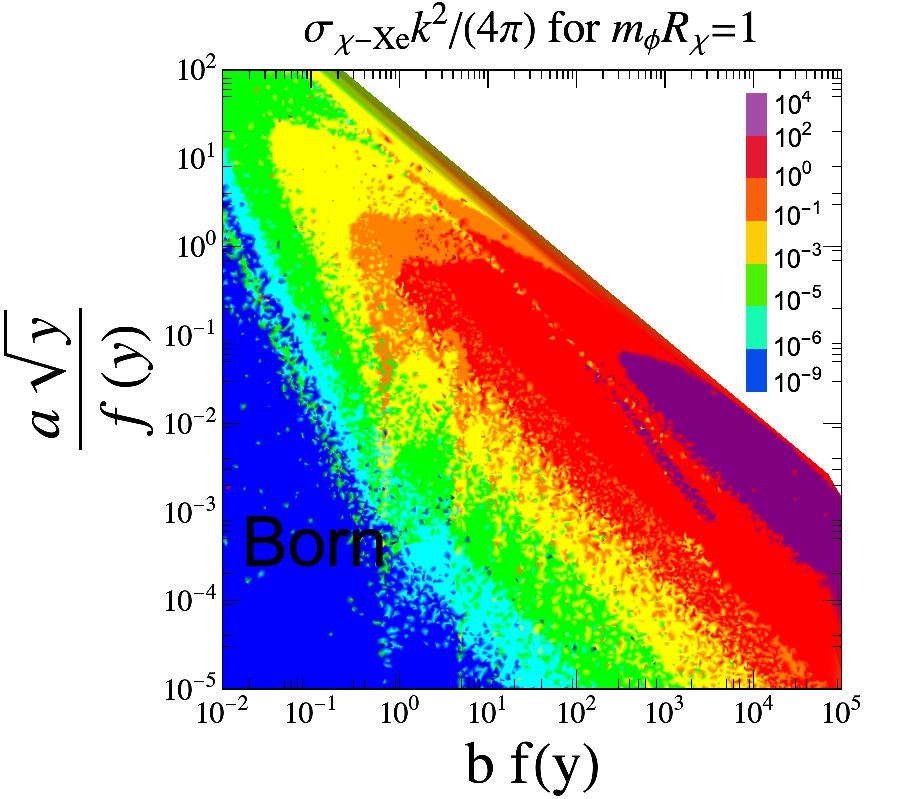}\hspace{-1mm}
       \vspace{-3mm}
\caption{Same as Fig. \ref{fig6}, but for  puffy dark matter-xenon nucleus scattering. }\label{fig7}
\end{figure}
Next, we study the scattering between a finite-size dark matter particle and a xenon nucleus. In this case, the potential function becomes
\begin{align}\label{eq23}
V_{\rm \chi-N}(r) & = \begin{cases}
~ g(r, 30.67n~\rm GeV^{-1}, 30.67~\rm GeV^{-1}) & r<2R_{\chi}\, , \\
\hspace{5cm}\  & \ \\[-6.mm]
~\alpha\frac{e^{-m_{\phi}r}}{r} 
\times h(r, 30.67n~\rm GeV^{-1}, 30.67~\rm GeV^{-1})&  r>2R_{\chi}\,.
\end{cases}
\end{align}
Similarly, by substituting the potential from Eq.(\ref{eq23}) into the Schrödinger equation Eq.(\ref{eq20}), we can obtain the scattering cross section. The parameter $bf(y)$ can be calculated by applying the Born approximation condition, substituting the potential from Eq.(\ref{eq23}) into Eq.(\ref{eq9}). Scanning the parameter space ($m_{\phi}$, $\alpha$, $M_{\chi}$) yields the results shown in Fig.~\ref{fig7} which is similar to Fig.~\ref{fig6}.  This figure indicates that with a large number of nucleons in the target nucleus, the scattering cross section between puffy dark matter and the nucleus can still be categorized into Born, resonance, and classical regions. Compared to the case with nucleon number equal to 1, for the same $y$ values the resonance peaks have larger values of the vertical parameter $a\sqrt{y}/f(y)$.

\subsection{Detection of nugget-type dark matter}\label{sec3.3}
The finite-size effect of particles can significantly impact their scattering cross section. For the scattering of large-size particles, as discussed in the preceding section, the cross section lies within the Born approximation regime. This implies that the interaction is effectively confined within the internal structure of the particles, resembling a short-range force. In this regime, the potential energy is much smaller than the kinetic energy, and thus the quantum mechanical and quantum field treatments are effectively equivalent. In contrast, for the scattering of small-size particles, this equivalence no longer holds, and accurate cross sections must be computed using partial wave analysis.  Beyond that, these results can also be seen in Figs.~5-7. The scattering cross sections between small-size particles occur in the resonance and classical regions. As for the Bonn region, it corresponds to the case of large-size superheavy dark matter particles studied in the literature using quantum field theory to calculate the scattering cross sections~\cite{Cappiello:2020lbk,Boukhtouchen:2025vvg,Clark:2020mna,Carney:2022gse,Aggarwal:2024ngx,LZ:2024psa}. Similarly, current research on nugget type dark matter focuses on the supermassive mass range~\cite{Wise:2014ola,Wise:2014jva,Gresham:2017cvl,Bai:2018dxf}. Our work, focusing on the direct detection of bound state dark matter composed of a small number of particles, is a supplement to the direct detection research of nugget type dark matter.  In the following we focus on the scattering between a finite-size bound state dark matter particle composed of a small number of constituent particles, such as a nugget-type dark matter particle with a small $N$, and a finite-size nucleus. Ref.~\cite{Gresham:2017zqi} utilizes relativistic mean field theory and gives a systematic computation of nugget properties.  For nugget dark matter with a small $N$, the constituent density is low and the constituents are non-relativistic. In this case, the self-interaction force range between dark matter particles is large (with the dark matter particle radius smaller than the force range, $R_{\chi}\leq m_{\phi}^{-1}$).  As a result, the effect of the mediator particle mass is not significant. The mass of the dark matter particle is approximately $Nm_{\chi}$.  Using the non-relativistic formula for a fermi gas, the radius of the dark matter particle can be estimated as $R_{\chi}\simeq[81\pi^{2}/(4Ng_{\rm dof}^{2}\alpha_{\chi}^{3}m_{\chi}^{3})]^{1/3}$ (see Ref.~\cite{Wise:2014jva} for details). Here, $\alpha_{\chi}$ is the coupling constant for the self-interaction of the dark matter particles, $g_{\rm dof}$ is the number of degrees of freedom for the fermion field. The stability condition of the bound state provides a relation between the mass and radius of the nugget dark matter particle. Furthermore, in the detection of nugget-type dark matter, the use of partial wave analysis allows for a more predictive parameter space for dark matter–nucleus scattering.

\begin{figure}[ht]
\centering
\includegraphics[width=7.2cm]{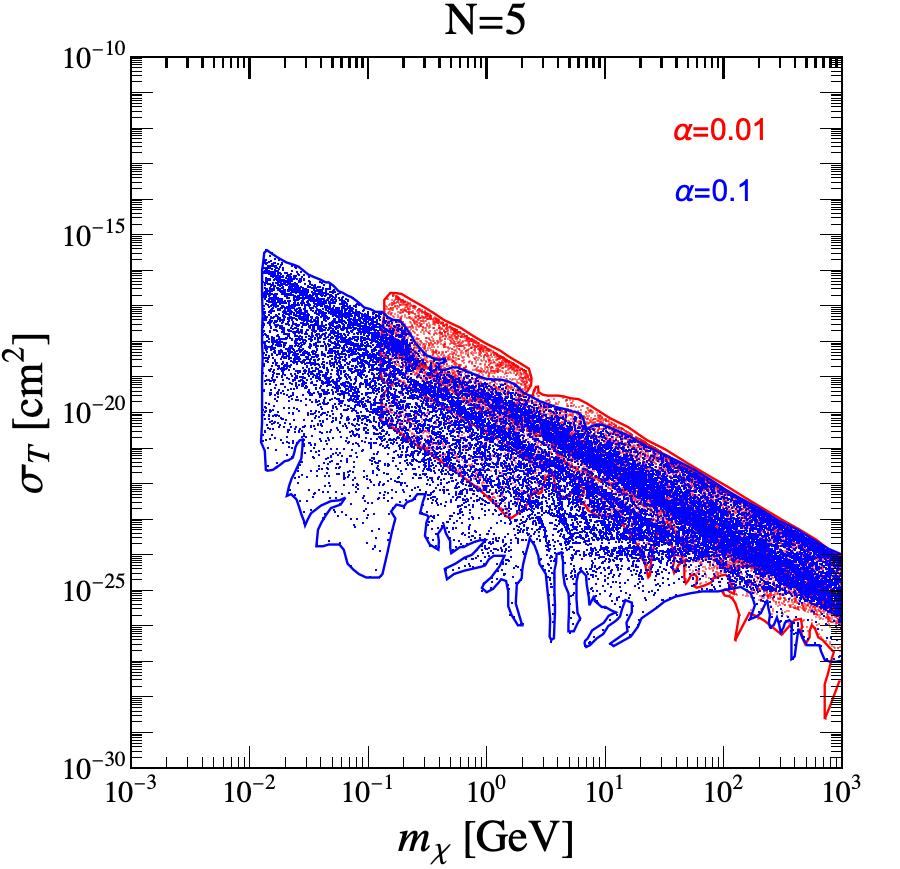}\hspace{-1mm}
\includegraphics[width=7.2cm]{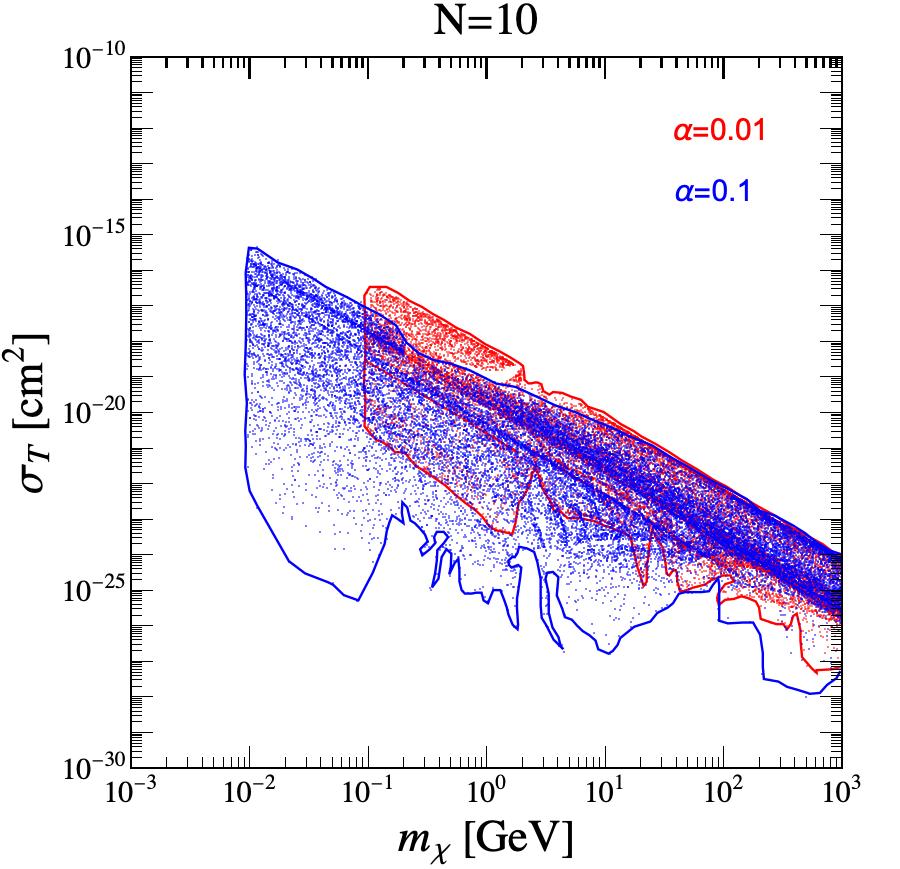}\hspace{-1mm}
     \vspace{-3mm}
\caption{The parameter space ($m_{\chi}, \sigma_T$) for nugget particle-Xe nucleus scattering  with a fixed coupling $\alpha=0.01$ (red points), 0.1 (blue points). The left panel corredponds to $\rm N=5$ and  the right panel to $\rm N=10$.}\label{fig8}
\end{figure}

Next, we consider a bound-state dark matter particle with constituent number $N=5$ or $N=10$. Given the self-coupling constant $\alpha_{\chi}=\alpha=0.01 (0.1)$, we substitute the potential function Eq.(\ref{eq23}) into the Schrödinger equation Eq.(\ref{eq20}). By scanning the parameter space $(m_{\phi}, m_{\chi})$, we obtain the nugget dark matter–xenon nucleus scattering cross section. As shown in Fig.~\ref{fig8}, the nugget dark matter–xenon nucleus scattering cross section does not exceed $10^{-16}\rm cm^2$ for $N=5$ and $N=10$.
The similar contour shapes in the left and right panels indicate that a small number of bound-state constituents has little effect on the nugget dark matter–nucleus scattering cross section. Moreover, the larger the coupling constant, the larger the allowed dark matter parameter space. For $\alpha=0.01 (0.1)$, the stability condition of the nugget-type bound state requires the dark matter particle mass to be greater than $0.1 (0.01)~\rm GeV$. In addition, one can also observe the appearance of resonance structures in the cross section from both panels.

Therefore, the finite size effects of particles in dark matter direct detection cannot be neglected. Even in the detection of point-like dark matter, the finite size of the target nucleus can significantly influence the scattering dynamics.  For the direct detection of finite-size dark matter, the scattering between particles of unequal sizes still allows for the classification into Born, resonant, and classical regimes. However, due to the diversity of the behavior of the interaction potential resulting from different particle sizes, the scattering dynamics become more complex. In the case of nugget-type dark matter particle, the stability condition of the bound-state provides the viable parameter space.

\section{Conclusion}\label{sec4}
Unlike the approach based on non-relativistic effective operators, we directly constructed the interaction potential between finite-size particles in coordinate space and then solved the Schrödinger equation to obtain the scattering cross section between non-relativistic finite-size particles. We found that due to the finite size effect of particles, the interaction potential between finite-size particles no longer exhibits the singular behavior of the point-particle Yukawa potential; instead, it approaches a constant value at the origin. The interaction potential between particles of unequal sizes was found to have a more complex dynamics. Applying this method to the study of dark matter direct detection, we obtained the following results: (i) For a point-like dark matter particle and a finite-size target nucleus, our re-calculation results agree with Ref.~\cite{Xu:2020qjk}, which shows the dark matter–nucleus scattering cross section can be classified into Born, resonant, and classical regimes; (ii) For the scattering between a puffy dark matter particle with a large size-to-range ratio and a finite-size nucleus, the cross section lies entirely within the Born regime. Since in this case the potential energy between the particles is much smaller than the kinetic energy, the results agree with those obtained from quantum field theory at the tree level; (iii) For a puffy dark matter particle with a small size-to-range ratio, the complex dynamics of the interaction potential gives rise to rich phenomenology in the scattering cross section between dark matter and finite-size nucleus, which can also be classified into Born, resonant, and classical regimes; (iv) For the direct detection of nugget-type dark matter with a small number of constituent particles, due to the stability condition of the dark matter bound-state,  the viable dark matter–nucleus scattering cross section region is specified.

\addcontentsline{toc}{section}{Acknowledgments}
\acknowledgments
We thank Bin Zhu for useful discussions. This work was supported by a Talent program from Chengdu Technological University (2024RC031), by the National Natural Science Foundation of China (NNSFC) under grant No. 12075300 and by a PI Research Fund from Henan Normal University (5101029470335). 

\appendix
\section{The  potential function}\label{appa}
The interaction potential between two finite particles is 
\begin{align}\label{eqa}
V_{\rm \chi-N}(r) & = \begin{cases}
~ g(r,R_{\chi},R_N) & r<2R_{\chi}\, , \\
\hspace{5cm}\  & \ \\[-6.mm]
~\alpha\frac{e^{-m_{\phi}r}}{r} 
\times h\left(R_{\chi},R_N\right)&  r>2R_{\chi}\,,
\end{cases}
\end{align}
where
\begin{eqnarray} 
\label{inbol}
g(r,R_{\chi},R_N)&=&\frac{3\alpha}{16\pi m_{\phi}^6R^3_{\chi}R^3_N}\left\{m^4_{\phi}(r-2R_{\chi})^2(r+4R_{\chi})
+\frac{6}{r}\left[-e^{m-{\phi}(r+R_{\chi}+R_N)}
\right.\right. \nonumber \\
&& \times (-2-2m_{\phi}R_{\chi} 
+e^{m_{\phi}R_{\chi}}(2+m_{\phi}R_{\chi}(-2+m_{\phi}(r-2R_{\chi}))+2m_{\phi}R_{\chi}))\nonumber\\
&& \times(1+m_{\phi}R_N+e^{2m_{\phi}R_{N}}(-1+m_{\phi}R_{N})) \nonumber\\
&& +e^{-m_{\phi}R_{N}}(1+m_{\phi}R_{N})(2(2+m^2_{\phi}r(r-2R_{\chi}))\cosh(m_{\phi}R_{\chi}) \nonumber \\
&& -4(\cosh[m_{\phi}(r-R_{\chi})]+m_{\phi}R_{\chi}\sinh[m_{\phi}(r-R_{\chi})])\nonumber\\
&& \left.\left. +4m_{\phi}(r-R_{\chi})\sinh(m_{\phi}R_{\chi})\right]\right\} 
\end{eqnarray} 
\begin{eqnarray} \label{outboll}
h\left(R_{\chi},R_N\right)&=&4\pi\left(\frac{3}{4\pi R_{\chi}^{3}}\right)\left(\frac{3}{4\pi R_{N}^{3}}\right)\frac{\pi}{m_{\phi}^{2}}\left(\frac{e^{-m_{\phi}R_{\chi}}}{m_{\phi}^{2}}-\frac{e^{m_{\phi}R_{\chi}}}{m_{\phi}^{2}}+\frac{R_{\chi}e^{-m_{\phi}R_{\chi}}}{m_{\phi}}+\frac{R_{\chi}e^{m_{\phi}R_{\chi}}}{m_{\phi}}\right)\nonumber \\
&&\times\left(\frac{e^{-m_{\phi}R_{N}}}{m_{\phi}^{2}}-\frac{e^{m_{\phi}R_{N}}}{m_{\phi}^{2}}+\frac{R_{N}e^{-m_{\phi}R_{N}}}{m_{\phi}}+\frac{R_{N}e^{m_{\phi}R_{N}}}{m_{\phi}}\right)
\end{eqnarray} 

\addcontentsline{toc}{section}{References}
\bibliographystyle{JHEP}
\bibliography{note}

@article{LZ:2024psa,
	author = "Aalbers, J. and others",
	collaboration = "LZ",
	title = "{New constraints on ultraheavy dark matter from the LZ experiment}",
	eprint = "2402.08865",
	archivePrefix = "arXiv",
	primaryClass = "hep-ex",
	reportNumber = "FERMILAB-PUB-24-0015-TD",
	doi = "10.1103/PhysRevD.109.112010",
	journal = "Phys. Rev. D",
	volume = "109",
	number = "11",
	pages = "112010",
	year = "2024"
}

@article{Aggarwal:2024ngx,
	author = "Aggarwal, Harsh and Raj, Nirmal",
	title = "{Ultraheavy multiscattering dark matter: DUNE, CYGNUS, other kilotonne detectors, and tidal streams}",
	eprint = "2410.22168",
	archivePrefix = "arXiv",
	primaryClass = "hep-ph",
	doi = "10.1103/PhysRevD.111.043010",
	journal = "Phys. Rev. D",
	volume = "111",
	number = "4",
	pages = "043010",
	year = "2025"
}

@article{Carney:2022gse,
	author = "Carney, Daniel and others",
	title = "{Snowmass2021 cosmic frontier white paper: Ultraheavy particle dark matter}",
	eprint = "2203.06508",
	archivePrefix = "arXiv",
	primaryClass = "hep-ph",
	reportNumber = "FERMILAB-PUB-22-160-T",
	doi = "10.21468/SciPostPhysCore.6.4.075",
	journal = "SciPost Phys. Core",
	volume = "6",
	pages = "075",
	year = "2023"
}

@article{Clark:2020mna,
	author = "Clark, Michael and Depoian, Amanda and Elshimy, Bahaa and Kopec, Abigail and Lang, Rafael F. and Li, Shengchao and Qin, Juehang",
	title = "{Direct Detection Limits on Heavy Dark Matter}",
	eprint = "2009.07909",
	archivePrefix = "arXiv",
	primaryClass = "hep-ph",
	doi = "10.1103/PhysRevD.102.123026",
	journal = "Phys. Rev. D",
	volume = "102",
	number = "12",
	pages = "123026",
	year = "2020"
}

@article{Boukhtouchen:2025vvg,
	author = "Boukhtouchen, Yilda and Bramante, Joseph and Cappiello, Christopher and Diamond, Melissa",
	title = "{Deconstructive Composite Dark Matter Detection}",
	eprint = "2512.16043",
	archivePrefix = "arXiv",
	primaryClass = "hep-ph",
	month = "12",
	year = "2025"
}

@article{Gelmini:2002ez,
	author = "Gelmini, Graciela and Kusenko, Alexander and Nussinov, Shmuel",
	title = "{Experimental identification of nonpointlike dark matter candidates}",
	eprint = "hep-ph/0203179",
	archivePrefix = "arXiv",
	reportNumber = "UCLA-01-TEP-7",
	doi = "10.1103/PhysRevLett.89.101302",
	journal = "Phys. Rev. Lett.",
	volume = "89",
	pages = "101302",
	year = "2002"
}

@article{Bai:2018dxf,
	author = "Bai, Yang and Long, Andrew J. and Lu, Sida",
	title = "{Dark Quark Nuggets}",
	eprint = "1810.04360",
	archivePrefix = "arXiv",
	primaryClass = "hep-ph",
	reportNumber = "FERMILAB-PUB-18-600-T",
	doi = "10.1103/PhysRevD.99.055047",
	journal = "Phys. Rev. D",
	volume = "99",
	number = "5",
	pages = "055047",
	year = "2019"
}

@article{Gresham:2017cvl,
	author = "Gresham, Moira I. and Lou, Hou Keong and Zurek, Kathryn M.",
	title = "{Early Universe synthesis of asymmetric dark matter nuggets}",
	eprint = "1707.02316",
	archivePrefix = "arXiv",
	primaryClass = "hep-ph",
	doi = "10.1103/PhysRevD.97.036003",
	journal = "Phys. Rev. D",
	volume = "97",
	number = "3",
	pages = "036003",
	year = "2018"
}

@article{Wise:2014jva,
	author = "Wise, Mark B. and Zhang, Yue",
	title = "{Stable Bound States of Asymmetric Dark Matter}",
	eprint = "1407.4121",
	archivePrefix = "arXiv",
	primaryClass = "hep-ph",
	reportNumber = "CALT-TH-2014-145",
	doi = "10.1103/PhysRevD.90.055030",
	journal = "Phys. Rev. D",
	volume = "90",
	number = "5",
	pages = "055030",
	year = "2014",
	note = "[Erratum: Phys.Rev.D 91, 039907 (2015)]"
}

@article{Wise:2014ola,
	author = "Wise, Mark B. and Zhang, Yue",
	title = "{Yukawa Bound States of a Large Number of Fermions}",
	eprint = "1411.1772",
	archivePrefix = "arXiv",
	primaryClass = "hep-ph",
	reportNumber = "CALT-TH-2014-160",
	doi = "10.1007/JHEP02(2015)023",
	journal = "JHEP",
	volume = "02",
	pages = "023",
	year = "2015",
	note = "[Erratum: JHEP 10, 165 (2015)]"
}

@article{Grabowska:2018lnd,
	author = "Grabowska, Dorota M. and Melia, Tom and Rajendran, Surjeet",
	title = "{Detecting Dark Blobs}",
	eprint = "1807.03788",
	archivePrefix = "arXiv",
	primaryClass = "hep-ph",
	doi = "10.1103/PhysRevD.98.115020",
	journal = "Phys. Rev. D",
	volume = "98",
	number = "11",
	pages = "115020",
	year = "2018"
}

@article{Colquhoun:2020adl,
	author = "Colquhoun, Brian and Heeba, Saniya and Kahlhoefer, Felix and Sagunski, Laura and Tulin, Sean",
	title = "{Semiclassical regime for dark matter self-interactions}",
	eprint = "2011.04679",
	archivePrefix = "arXiv",
	primaryClass = "hep-ph",
	reportNumber = "TTK-20-39",
	doi = "10.1103/PhysRevD.103.035006",
	journal = "Phys. Rev. D",
	volume = "103",
	number = "3",
	pages = "035006",
	year = "2021"
}

@article{Cappiello:2020lbk,
	author = "Cappiello, Christopher V. and Collar, J. I. and Beacom, John F.",
	title = "{New experimental constraints in a new landscape for composite dark matter}",
	eprint = "2008.10646",
	archivePrefix = "arXiv",
	primaryClass = "hep-ex",
	doi = "10.1103/PhysRevD.103.023019",
	journal = "Phys. Rev. D",
	volume = "103",
	number = "2",
	pages = "023019",
	year = "2021"
}

@article{Bhoonah:2020dzs,
	author = "Bhoonah, Amit and Bramante, Joseph and Schon, Sarah and Song, Ningqiang",
	title = "{Detecting composite dark matter with long-range and contact interactions in gas clouds}",
	eprint = "2010.07240",
	archivePrefix = "arXiv",
	primaryClass = "hep-ph",
	doi = "10.1103/PhysRevD.103.123026",
	journal = "Phys. Rev. D",
	volume = "103",
	number = "12",
	pages = "123026",
	year = "2021"
}

@article{Wang:2021tjf,
	author = "Wang, Wenyu and Xu, Wu-Long and Zhu, Bin",
	title = "{Realistic scattering of puffy dark matter}",
	eprint = "2108.07030",
	archivePrefix = "arXiv",
	primaryClass = "hep-ph",
	doi = "10.1103/PhysRevD.105.075013",
	journal = "Phys. Rev. D",
	volume = "105",
	number = "7",
	pages = "075013",
	year = "2022"
}

@article{Digman:2019wdm,
	author = "Digman, Matthew C. and Cappiello, Christopher V. and Beacom, John F. and Hirata, Christopher M. and Peter, Annika H. G.",
	title = "{Not as big as a barn: Upper bounds on dark matter-nucleus cross sections}",
	eprint = "1907.10618",
	archivePrefix = "arXiv",
	primaryClass = "hep-ph",
	doi = "10.1103/PhysRevD.100.063013",
	journal = "Phys. Rev. D",
	volume = "100",
	number = "6",
	pages = "063013",
	year = "2019",
	note = "[Erratum: Phys.Rev.D 106, 089902 (2022)]"
}

@article{Beneke:2022rjv,
	author = "Beneke, Martin and Lederer, Stefan and Urban, Kai",
	title = "{Sommerfeld enhancement of resonant dark matter annihilation}",
	eprint = "2209.14343",
	archivePrefix = "arXiv",
	primaryClass = "hep-ph",
	reportNumber = "TUM-HEP-1419/22",
	doi = "10.1016/j.physletb.2023.137773",
	journal = "Phys. Lett. B",
	volume = "839",
	pages = "137773",
	year = "2023"
}

@phdthesis{Bollig:2024ipe,
	author = "Bollig, Julian",
	title = "{The impact of non-perturbative effects in dark matter production and detection}",
	eprint = "2409.14813",
	archivePrefix = "arXiv",
	primaryClass = "hep-ph",
	doi = "10.6094/UNIFR/255791",
	school = "Freiburg U.",
	year = "2024"
}

@article{Bollig:2021psb,
	author = "Bollig, Julian and Vogl, Stefan",
	title = "{Impact of bound states on non-thermal dark matter production}",
	eprint = "2112.01491",
	archivePrefix = "arXiv",
	primaryClass = "hep-ph",
	doi = "10.1088/1475-7516/2022/10/031",
	journal = "JCAP",
	volume = "10",
	pages = "031",
	year = "2022"
}

@article{Hardy:2015boa,
	author = "Hardy, Edward and Lasenby, Robert and March-Russell, John and West, Stephen M.",
	title = "{Signatures of Large Composite Dark Matter States}",
	eprint = "1504.05419",
	archivePrefix = "arXiv",
	primaryClass = "hep-ph",
	doi = "10.1007/JHEP07(2015)133",
	journal = "JHEP",
	volume = "07",
	pages = "133",
	year = "2015"
}

@article{Acevedo:2021kly,
	author = "Acevedo, Javier F. and Bramante, Joseph and Goodman, Alan",
	title = "{Accelerating composite dark matter discovery with nuclear recoils and the Migdal effect}",
	eprint = "2108.10889",
	archivePrefix = "arXiv",
	primaryClass = "hep-ph",
	doi = "10.1103/PhysRevD.105.023012",
	journal = "Phys. Rev. D",
	volume = "105",
	number = "2",
	pages = "023012",
	year = "2022"
}

@article{Kurinsky:2019pgb,
	author = "Kurinsky, Noah Alexander and Yu, To Chin and Hochberg, Yonit and Cabrera, Blas",
	title = "{Diamond Detectors for Direct Detection of Sub-GeV Dark Matter}",
	eprint = "1901.07569",
	archivePrefix = "arXiv",
	primaryClass = "hep-ex",
	reportNumber = "FERMILAB-PUB-19-020-AE-E",
	doi = "10.1103/PhysRevD.99.123005",
	journal = "Phys. Rev. D",
	volume = "99",
	number = "12",
	pages = "123005",
	year = "2019"
}

@article{Hochberg:2016ajh,
	author = "Hochberg, Yonit and Lin, Tongyan and Zurek, Kathryn M.",
	title = "{Detecting Ultralight Bosonic Dark Matter via Absorption in Superconductors}",
	eprint = "1604.06800",
	archivePrefix = "arXiv",
	primaryClass = "hep-ph",
	doi = "10.1103/PhysRevD.94.015019",
	journal = "Phys. Rev. D",
	volume = "94",
	number = "1",
	pages = "015019",
	year = "2016"
}

@article{Liang:2022xbu,
	author = "Liang, Zheng-Liang and Mo, Chongjie and Zheng, Fawei and Zhang, Ping",
	title = "{Phonon-mediated Migdal effect in semiconductor detectors}",
	eprint = "2205.03395",
	archivePrefix = "arXiv",
	primaryClass = "hep-ph",
	doi = "10.1103/PhysRevD.106.043004",
	journal = "Phys. Rev. D",
	volume = "106",
	number = "4",
	pages = "043004",
	year = "2022",
	note = "[Erratum: Phys.Rev.D 106, 109901 (2022)]"
}

@article{Cheng:2023loy,
	author = "Cheng, Xi and Guo, Ji-Heng and Wang, Wenyu and Zhu, Bin",
	title = "{Probing levitodynamics with multi-stochastic forces and the simple applications on the dark matter detection in optical levitation experiment}",
	eprint = "2312.04202",
	archivePrefix = "arXiv",
	primaryClass = "cond-mat.stat-mech",
	doi = "10.1016/j.nuclphysb.2024.116780",
	journal = "Nucl. Phys. B",
	volume = "1010",
	pages = "116780",
	year = "2025"
}

@article{Kilian:2024fsg,
	author = "Kilian, Eva and others",
	title = "{Dark Matter Searches with Levitated Sensors}",
	eprint = "2401.17990",
	archivePrefix = "arXiv",
	primaryClass = "quant-ph",
	doi = "10.1116/5.0200916",
	journal = "AVS Quantum Sci.",
	volume = "6",
	pages = "030503",
	year = "2024"
}

@article{LZ:2025iaw,
	author = "Aalbers, J. and others",
	collaboration = "LZ",
	title = "{New Constraints on Cosmic Ray-Boosted Dark Matter from the LUX-ZEPLIN Experiment}",
	eprint = "2503.18158",
	archivePrefix = "arXiv",
	primaryClass = "hep-ex",
	reportNumber = "FERMILAB-PUB-25-0191-V",
	doi = "10.1103/nr92-jvt3",
	journal = "Phys. Rev. Lett.",
	volume = "134",
	number = "24",
	pages = "241801",
	year = "2025"
}

@article{Knapen:2020aky,
	author = "Knapen, Simon and Kozaczuk, Jonathan and Lin, Tongyan",
	title = "{Migdal Effect in Semiconductors}",
	eprint = "2011.09496",
	archivePrefix = "arXiv",
	primaryClass = "hep-ph",
	doi = "10.1103/PhysRevLett.127.081805",
	journal = "Phys. Rev. Lett.",
	volume = "127",
	number = "8",
	pages = "081805",
	year = "2021"
}

@article{XENON:2024znc,
	author = "Aprile, E. and others",
	collaboration = "XENON, (XENON Collaboration)**",
	title = "{Search for Light Dark Matter in Low-Energy Ionization Signals from XENONnT}",
	eprint = "2411.15289",
	archivePrefix = "arXiv",
	primaryClass = "hep-ex",
	doi = "10.1103/PhysRevLett.134.161004",
	journal = "Phys. Rev. Lett.",
	volume = "134",
	number = "16",
	pages = "161004",
	year = "2025"
}

@article{Essig:2017kqs,
	author = "Essig, Rouven and Volansky, Tomer and Yu, Tien-Tien",
	title = "{New Constraints and Prospects for sub-GeV Dark Matter Scattering off Electrons in Xenon}",
	eprint = "1703.00910",
	archivePrefix = "arXiv",
	primaryClass = "hep-ph",
	reportNumber = "CERN-TH-2017-042, YITP-SB-17-09",
	doi = "10.1103/PhysRevD.96.043017",
	journal = "Phys. Rev. D",
	volume = "96",
	number = "4",
	pages = "043017",
	year = "2017"
}

@article{XENON:2025vwd,
	author = "Aprile, E. and others",
	collaboration = "XENON",
	title = "{WIMP Dark Matter Search using a 3.1 tonne $\times$ year Exposure of the XENONnT Experiment}",
	eprint = "2502.18005",
	archivePrefix = "arXiv",
	primaryClass = "hep-ex",
	month = "2",
	year = "2025"
}

@article{XENON:2024wpa,
	author = "Aprile, E. and others",
	collaboration = "XENON",
	title = "{The XENONnT dark matter experiment}",
	eprint = "2402.10446",
	archivePrefix = "arXiv",
	primaryClass = "physics.ins-det",
	doi = "10.1140/epjc/s10052-024-12982-5",
	journal = "Eur. Phys. J. C",
	volume = "84",
	number = "8",
	pages = "784",
	year = "2024"
}

@article{GAMBIT:2017zdo,
	author = "Athron, Peter and others",
	collaboration = "GAMBIT",
	title = "{A global fit of the MSSM with GAMBIT}",
	eprint = "1705.07917",
	archivePrefix = "arXiv",
	primaryClass = "hep-ph",
	reportNumber = "COEPP-MN-17-11, NORDITA-2017-081, CERN-TH-2017-169, CoEPP-MN-17-11, NORDITA 2017-081, gambit-physics, NORDITA 2017-081,
	gambit-physics-2017",
	doi = "10.1140/epjc/s10052-017-5196-8",
	journal = "Eur. Phys. J. C",
	volume = "77",
	number = "12",
	pages = "879",
	year = "2017"
}

@article{Lee:1977ua,
	author = "Lee, Benjamin W. and Weinberg, Steven",
	editor = "Srednicki, M. A.",
	title = "{Cosmological Lower Bound on Heavy Neutrino Masses}",
	reportNumber = "FERMILAB-PUB-77-041-T",
	doi = "10.1103/PhysRevLett.39.165",
	journal = "Phys. Rev. Lett.",
	volume = "39",
	pages = "165--168",
	year = "1977"
}

@article{Jungman:1995df,
	author = "Jungman, Gerard and Kamionkowski, Marc and Griest, Kim",
	title = "{Supersymmetric dark matter}",
	eprint = "hep-ph/9506380",
	archivePrefix = "arXiv",
	reportNumber = "SU-4240-605, UCSD-PTH-95-02, IASSNS-HEP-95-14, CU-TP-677",
	doi = "10.1016/0370-1573(95)00058-5",
	journal = "Phys. Rept.",
	volume = "267",
	pages = "195--373",
	year = "1996"
}

@article{Buchmueller:2017qhf,
	author = "Buchmueller, Oliver and Doglioni, Caterina and Wang, Lian Tao",
	title = "{Search for dark matter at colliders}",
	eprint = "1912.12739",
	archivePrefix = "arXiv",
	primaryClass = "hep-ex",
	doi = "10.1038/nphys4054",
	journal = "Nature Phys.",
	volume = "13",
	number = "3",
	pages = "217--223",
	year = "2017"
}

@article{Lin:2019uvt,
	author = "Lin, Tongyan",
	title = "{Dark matter models and direct detection}",
	eprint = "1904.07915",
	archivePrefix = "arXiv",
	primaryClass = "hep-ph",
	doi = "10.22323/1.333.0009",
	journal = "PoS",
	volume = "333",
	pages = "009",
	year = "2019"
}

@article{Randall:2008ppe,
	author = "Randall, Scott W. and Markevitch, Maxim and Clowe, Douglas and Gonzalez, Anthony H. and Bradac, Marusa",
	title = "{Constraints on the Self-Interaction Cross-Section of Dark Matter from Numerical Simulations of the Merging Galaxy Cluster 1E 0657-56}",
	eprint = "0704.0261",
	archivePrefix = "arXiv",
	primaryClass = "astro-ph",
	doi = "10.1086/587859",
	journal = "Astrophys. J.",
	volume = "679",
	pages = "1173--1180",
	year = "2008"
}

@article{Ostriker:1973uit,
	author = "Ostriker, J. P. and Peebles, P. J. E.",
	title = "{A Numerical Study of the Stability of Flattened Galaxies: or, can Cold Galaxies Survive?}",
	doi = "10.1086/152513",
	journal = "Astrophys. J.",
	volume = "186",
	pages = "467--480",
	year = "1973"
}

@article{Tian:2020tur,
	author = "Tian, S. X.",
	title = "{Cosmological consequences of a scalar field with oscillating equation of state. II. Oscillating scaling and chaotic accelerating solutions}",
	eprint = "2010.03314",
	archivePrefix = "arXiv",
	primaryClass = "gr-qc",
	doi = "10.1103/PhysRevD.102.063509",
	journal = "Phys. Rev. D",
	volume = "102",
	number = "6",
	pages = "063509",
	year = "2020"
}

@article{Bertone:2016nfn,
	author = "Bertone, Gianfranco and Hooper, Dan",
	title = "{History of dark matter}",
	eprint = "1605.04909",
	archivePrefix = "arXiv",
	primaryClass = "astro-ph.CO",
	reportNumber = "FERMILAB-PUB-16-157-A",
	doi = "10.1103/RevModPhys.90.045002",
	journal = "Rev. Mod. Phys.",
	volume = "90",
	number = "4",
	pages = "045002",
	year = "2018"
}

@article{Gresham:2017zqi,
	author = "Gresham, Moira I. and Lou, Hou Keong and Zurek, Kathryn M.",
	title = "{Nuclear Structure of Bound States of Asymmetric Dark Matter}",
	eprint = "1707.02313",
	archivePrefix = "arXiv",
	primaryClass = "hep-ph",
	doi = "10.1103/PhysRevD.96.096012",
	journal = "Phys. Rev. D",
	volume = "96",
	number = "9",
	pages = "096012",
	year = "2017"
}

@article{Chu:2018faw,
	author = "Chu, Xiaoyong and Garcia-Cely, Camilo and Murayama, Hitoshi",
	title = "{Finite-size dark matter and its effect on small-scale structure}",
	eprint = "1901.00075",
	archivePrefix = "arXiv",
	primaryClass = "hep-ph",
	reportNumber = "DESY-18-225, IPMU18-0207",
	doi = "10.1103/PhysRevLett.124.041101",
	journal = "Phys. Rev. Lett.",
	volume = "124",
	number = "4",
	pages = "041101",
	year = "2020"
}

@article{Wang:2023wrx,
	author = "Wang, Wenyu and Xu, Wu-Long and Yang, Jin Min and Zhu, Rui",
	title = "{Direct detection of cosmic ray-boosted puffy dark matter}",
	eprint = "2305.12668",
	archivePrefix = "arXiv",
	primaryClass = "hep-ph",
	doi = "10.1016/j.nuclphysb.2023.116348",
	journal = "Nucl. Phys. B",
	volume = "995",
	pages = "116348",
	year = "2023"
}

@article{Wang:2023xgm,
	author = "Wang, Wenyu and Xu, Wu-Long and Yang, Jin Min",
	title = "{Direct detection of finite-size dark matter via electron recoil}",
	eprint = "2304.13243",
	archivePrefix = "arXiv",
	primaryClass = "hep-ph",
	doi = "10.1142/S0217751X23501440",
	journal = "Int. J. Mod. Phys. A",
	volume = "38",
	number = "26n27",
	pages = "2350144",
	year = "2023"
}

@article{Tulin:2013teo,
	author = "Tulin, Sean and Yu, Hai-Bo and Zurek, Kathryn M.",
	title = "{Beyond Collisionless Dark Matter: Particle Physics Dynamics for Dark Matter Halo Structure}",
	eprint = "1302.3898",
	archivePrefix = "arXiv",
	primaryClass = "hep-ph",
	doi = "10.1103/PhysRevD.87.115007",
	journal = "Phys. Rev. D",
	volume = "87",
	number = "11",
	pages = "115007",
	year = "2013"
}

@article{Xu:2020qjk,
	author = "Xu, Xingchen and Farrar, Glennys R.",
	title = "{Resonant scattering between dark matter and baryons: Revised direct detection and CMB limits}",
	eprint = "2101.00142",
	archivePrefix = "arXiv",
	primaryClass = "hep-ph",
	doi = "10.1103/PhysRevD.107.095028",
	journal = "Phys. Rev. D",
	volume = "107",
	number = "9",
	pages = "095028",
	year = "2023"
}

@article{An:2024wmc,
    author = "An, Haipeng and Ge, Shuailiang and Liu, Jia and Liu, Mingzhe",
    title = "{In Situ Measurements of Dark Photon Dark Matter Using Parker Solar Probe: Going beyond the Radio Window}",
    eprint = "2405.12285",
    archivePrefix = "arXiv",
    primaryClass = "hep-ph",
    doi = "10.1103/PhysRevLett.134.171001",
    journal = "Phys. Rev. Lett.",
    volume = "134",
    number = "17",
    pages = "171001",
    year = "2025"
}

@article{Liang:2024xcx,
	author = "Liang, Zheng-Liang and Su, Liangliang and Wu, Lei and Zhu, Bin",
	title = "{Plasmon-enhanced Direct Detection of sub-MeV Dark Matter}",
	eprint = "2401.11971",
	archivePrefix = "arXiv",
	primaryClass = "hep-ph",
	doi = "10.1103/PhysRevLett.134.071001",
	journal = "Phys. Rev. Lett.",
	volume = "134",
	number = "7",
	pages = "071001",
	year = "2025"
}

@article{Wang:2023xii,
	author = "Wang, Wenyu and Xu, Wu-Long and Yang, Jin Min and Zhu, Bin",
	title = "{Revisiting puffy dark matter with novel insights: partial wave analysis}",
	eprint = "2303.11058",
	archivePrefix = "arXiv",
	primaryClass = "hep-ph",
	doi = "10.1007/JHEP06(2023)103",
	journal = "JHEP",
	volume = "06",
	pages = "103",
	year = "2023"
}

@article{Wang:2021jic,
	author = "Wang, Jin-Wei and Granelli, Alessandro and Ullio, Piero",
	title = "{Direct Detection Constraints on Blazar-Boosted Dark Matter}",
	eprint = "2111.13644",
	archivePrefix = "arXiv",
	primaryClass = "astro-ph.HE",
	doi = "10.1103/PhysRevLett.128.221104",
	journal = "Phys. Rev. Lett.",
	volume = "128",
	number = "22",
	pages = "221104",
	year = "2022"
}

@article{Wang:2023wbw,
	author = "Wang, Wenyu and Xu, Wu-Long and Yang, Jin Min and Zhu, Bin and Zhu, Rui",
	title = "{Sommerfeld enhancement for puffy self-interacting dark matter}",
	eprint = "2308.02170",
	archivePrefix = "arXiv",
	primaryClass = "hep-ph",
	doi = "10.1007/JHEP01(2024)114",
	journal = "JHEP",
	volume = "01",
	pages = "114",
	year = "2024"
}

\end{document}